\documentclass{article}

\usepackage{PRIMEarxiv}

\usepackage[utf8]{inputenc} % allow utf-8 input
\usepackage[T1]{fontenc}    % use 8-bit T1 fonts
\usepackage{hyperref}       % hyperlinks
\usepackage{url}            % simple URL typesetting
\usepackage{booktabs}       % professional-quality tables
\usepackage{amsfonts}       % blackboard math symbols
\usepackage{nicefrac}       % compact symbols for 1/2, etc.
\usepackage{microtype}      % microtypography
\usepackage{lipsum}
\usepackage{fancyhdr}       % header
\usepackage{graphicx}       % graphics
\usepackage{amsmath}       % math

\graphicspath{{media/}}     % organize your images and other figures under media/ folder

\title{Bayesian modelling of the temporal evolution of seismicity using the \texttt{ETAS.inlabru} package
}

\author{
  Mark Naylor \\
  School of Geosciences \\
  University of Edinburgh \\
  \texttt{mark.naylor@ed.ac.uk} \\
  \And
  Francesco Serafini \\
  School of Geosciences \\
  University of Edinburgh \\
  \texttt{francesco.serafini@ed.ac.uk} \\
  \And 
  Finn Lindgren \\
  School of Mathematics \\
  University of Edinburgh \\
  \texttt{finn.lindgren@ed.ac.uk} \\
  \And
  Ian Main \\
  School of Geosciences \\
  University of Edinburgh \\
  \texttt{ian.main@ed.ac.uk} \\%% \AND
}

\newcommand{\inlabru}[0]{\texttt{inlabru}\,}
\newcommand{\bayesianETAS}[0]{\texttt{bayesianETAS}\, }

\begin{document}
\maketitle

\begin{abstract}
The Epidemic Type Aftershock Sequence (ETAS) model is widely used to model seismic sequences and underpins Operational Earthquake Forecasting (OEF). However, it remains challenging to assess the reliability of inverted ETAS parameters for a range of reasons. For example, the most common algorithms just return point estimates with little quantification of uncertainty. At the same time, Bayesian Markov Chain Monte Carlo implementations remain slow to run, do not scale well and few have been extended to include spatial structure. This makes it difficult to explore the effects of stochastic uncertainty. Here we present a new approach to ETAS modelling using an alternative Bayesian method, the Integrated Nested Laplace Approximation (INLA). We have implemented this model in a new R-Package called \texttt{ETAS.inlabru}, which builds on the R packages R-INLA and \inlabru. Our work has included extending these packages, which provided tools for modelling log-Gaussian Cox processes, to include the self-exciting Hawkes process that ETAS is a special case of. Whilst we just present the temporal component here, the model scales to a spatio-temporal model and may include a variety of spatial covariates. This is a fast method which returns joint posteriors on the ETAS background and triggering parameters. Using a series of synthetic case studies, we explore the robustness of ETAS inversions using this method of inversion. We also included runnable notebooks to reproduce the figures in this paper as part of the package's GitHub repository. We demonstrate that reliable estimates of the model parameters require that the catalogue data contains periods of relative quiescence as well as triggered sequences. We explore the robustness under stochastic uncertainty in the training data and show that the method is robust to a wide range of starting conditions. We show how the inclusion of historic earthquakes prior to the modelled domain affects the quality of the inversion. Finally, we show that rate dependent incompleteness after large earthquakes has a significant and detrimental effect on the ETAS posteriors. We believe that the speed of the \inlabru inversion, which include a rigorous estimation of uncertainty, will enable a deeper exploration of how to use ETAS robustly for seismicity modelling and operational earthquake forecasting.
\end{abstract}

% keywords can be removed

\section{Introduction}
The Epidemic Type Aftershock Sequence model (ETAS) (\cite{ogata1988etas, ogata2006space, ogata2011significant}) is one of the cornerstones of seismicity modelling. It models evolving seismic sequences in terms of background seismicity and seismicity triggered by previous events. As such, it is a self-exciting point process model which is commonly termed a Hawkes process (\cite{hawkes1971spectra}) in the statistical literature. ETAS achieves this by combining several empirical relationships for seismicity. The ETAS model enables us to generate synthetic earthquake sequences and to invert earthquake space-time-magnitude data for the underlying ETAS parameters that characterise both the background and triggering rates. However, the likelihood space for some parameters is notoriously flat and many factors can affect the robustness of the results.

There are many different implementations of the ETAS model. The most common approach for determining ETAS parameters is the maximum likelihood method which returns a point estimate on the ETAS parameters using an optimisation algorithm (e.g. \cite{JSSv088c01}). In some cases uncertainty is quoted using the Hessian matrix. Bayesian alternatives are available; for example the "\bayesianETAS" R-package (\cite{ross2021bayesian}) uses the Markov Chain Monte Carlo (MCMC) method to return full posteriors. However MCMC methods are notoriously slow as building the Markov Chain is an inherently linear algorithm requiring many successive runs of a forward model based on the previous results. A major benefit of Bayesian methods is that they better describe uncertainty. We have developed a new Bayesian ETAS package using the Integrated Nested Laplace Approximation (INLA) instead of MCMC; this is implemented in the R-Package \texttt{ETAS.inlabru} and will be made available through the Comprehensive R Archive Network (CRAN). The results presented in this paper are reproducible using this package and a series of Rmd notebooks. Unlike the MCMC implementations, our method does not rely on a latent variable to classify whether events are background or triggered.

The Integrated Nested Laplace Approximation (INLA, \cite{rue2017bayesian}) and \inlabru  \cite{bachl2019inlabru} offer a fast approach for Bayesian modelling of spatial, temporal and spatio-temporal point process data and have had over 10 years of development. The INLA method is a well-known alternative to MCMC methods to perform Bayesian inference. It has been successfully applied in a variety of fields such as seismic hazard (\cite{bayliss2020data,baylissCalifornia2022}), air pollution (\cite{forlani2020joint}), disease mapping (\cite{riebler2016intuitive, santermans2016spatiotemporal, schrodle2011primer, schrodle2011spatio}), genetics (\cite{opitz2016extensive}), public health (\cite{halonen2015road, }), ecology (\cite{roos2015modeling, teng2022bayesian}), more examples can be found in \cite{bakka2018spatial, blangiardo2013spatial, gomez2020bayesian}. 

To date, the main limitation for the application of \inlabru to seismicity was that it only addressed log-Gaussian Cox Processes \cite{taylor2014inla}, which do not include self-exciting clustering. \cite{serafini2022approximation} addressed this specific limitation by showing how the methodology used for log-Gaussian Cox processes could be extended to model self-exciting Hawkes Processes (\cite{hawkes1971point, hawkes1971spectra}), using R-INLA and \inlabru, when the function form of the triggering function can be integrated. The novelty of our approach resides in the likelihood approximation. We decompose the log-likelihood into the sum of many small components, where each is linearly approximated with respect to the posterior mode using a Taylor expansion. This means that the log-likelihood is exact at the posterior mode and the accuracy of the approximation decreases as we move away from that point. Furthermore, the linear approximation and the optimization routine to determine the posterior mode are internally performed by the \inlabru package. In this work, the specific application to the ETAS model was presented. The temporal model provides posteriors on the background rate and all ETAS parameters. 

\texttt{ETAS.inlabru} provides the functions to be approximated whilst the user provides the data and specifies the priors. The advantages of our approach are both in terms of computational time and its scalability to include relevant covariates (\cite{bayliss2020data}) such as maps of faults, strain rates etc. in addition to earthquake catalogue data, and/or to introduce alternative structures to the parameters (e.g. considering one of them as temporally, or spatially, varying).

Here, we present a broad analysis of how the \inlabru inversion performs on synthetic earthquake catalogues where we know all of the controlling parameters. We explore the performance of the inversion as a function of the training catalogue length, the impact of large events that happen to occur in the sequence, the consequence of short term incompleteness after large events as well as various \inlabru model choices. These results are generic and not specific to our implementation of the ETAS model - rather our fast Bayesian model allows us to make a more rapid assessment of potential biases derived from the likelihood function itself. We want the reader to come away with an understanding of when the ETAS model is likely to describe a sequence well, and to be able to identify sources of potential bias, understand how synthetic modelling allows us to explore potential data quality issues, and decide whether fitting an ETAS model is an appropriate way to proceed. We conclude with a demonstration of how the results can be used to develop a temporal Operational Earthquake Forecast.
\section{Method}

In this section, we introduce our \inlabru implementation of the temporal ETAS model and refer the reader to \cite{serafini2022approximation} for a complete description of the mathematical formulation.

\subsection{The temporal ETAS model}

The temporal ETAS model is a marked Hawkes process model, where the marking variable is the magnitude of the events. The ETAS model is composed of three parts: a background rate term, a triggered events rate representing the rate of events induced by past seismicity, and a magnitude distribution independent from space and time. Given the independence between the magnitude distribution and the time distribution of the events, the ETAS conditional intensity is usually the product between a Hawkes process model describing only the location and a magnitude distribution $\pi(m)$. 

More formally, the ETAS conditional intensity function evaluated at a generic time point $t \in (T_1, T_2), T_1, T_2 \geq 0, T_1 < T_2$ having observed the events $\mathcal H_t = \{(t_h, m_h) : t_h < t, m_h > M_0,\, \forall h = 1,..., N(t^{-})\}$, where $M_0$ is the minimum magnitude in the catalogue which needs to be completely sampled, and $N(t)$ is the counting process associated with the Hawkes process representing the number of events recorded up to time $t$ (included), is given by:

\begin{equation} \label{eq:lambda_ETAS}
    \lambda_{ETAS}(t, m | \mathcal H_t) = \lambda_{Hawkes}(t | \mathcal H_t)\pi(m) 
\end{equation}
where $\lambda_{Hawkes}$ is the conditional intensity of a temporal Hawkes process describing the occurrence times only. In our ETAS implementation this is given by:

\begin{equation} \label{eq:lambda_hawkes}
\lambda_{Hawkes}(t | \mathcal H_t) = \mu + \sum_{(t_h, m_h) \in \mathcal H_t} K e^{\alpha(m_h - M_0)}\left(\frac{t - t_h}{c} + 1\right)^{-p}    
\end{equation}

The parameters of the model are $\mu, K, \alpha, c \geq 0$ and $p > 1$. Different parametrisations of the ETAS model exist; we focus on this one because it has proven to be the most suitable parametrisation for our method. 

In seismology, the magnitude distribution, $\pi(m)$ is commonly assumed to be independent of space and time for simplicity of analysis. In this work, we take this to be the Gutenberg-Richter distribution with a $b$-value of 1. In this section, we focus on the Hawkes part of the model assuming the parameters of the magnitude distribution are determined independently. From now on, for ease of notation, where not specified differently we refer to $\lambda_{Hawkes}$ as simply $\lambda$.

\subsection{Hawkes Process Log-likelihood approximation for \inlabru}

The Hawkes process is implemented in \inlabru by decomposing its log-likelihood function (Eqn.\ref{eq:HP_LL}) into multiple parts, the sum of which returns the exact log-likelihood at the point we expand it about. We linearly approximate the single components with respect to the posterior mode and apply the \emph{Integrated Nested Laplace Approximation} (INLA) method to perform inference on the parameters of the model. Both the linearisation and the optimization, to find the posterior mode, are performed internally by \inlabru. Our package, \texttt{ETAS.inlabru}, provides \inlabru with the ETAS specific functions representing the log-likelihood components to be approximated. We outline the decomposition below.

Having observed a catalogue of events $\mathcal H = \{(t_i, m_i) : t_i \in [T_1,T_2], m_i \in (M_0, \infty)\}$, the Hawkes process log-likelihood is given by:

\begin{equation} \label{eq:HP_LL}
\mathcal L(\boldsymbol \theta | \mathcal H) = -\Lambda(T_1, T_2) + \sum_{(t_i, m_i) \in \mathcal H} \log\lambda(t_i | \mathcal H_{t_i}) 
\end{equation}

Where $\boldsymbol \theta$ is a vector of the model parameters, $\mathcal H_{t_i} = \{(t_h, m_h) \in \mathcal H : t_h < t_i\}$ is the history of events up to time $t_i$, and 

\begin{equation}
\begin{aligned}
\Lambda(T_1, T_2) & = \int_{T_1}^{T_2} \lambda(t|\mathcal H_t) dt \\
& = (T_2 - T_1)\mu + \sum_{(t_i, m_i) \in \mathcal H} \int_{T_1}^{T_2} K e^{\alpha(m_i - M_0)}\left(\frac{t - t_i}{c} + 1\right)^{-p} \mathbb I(t > t_i) dt \\
& = (T_2 - T_1)\mu + \sum_{(t_i, m_i)\in \mathcal H}K e^{\alpha(m_i - M_0)} \int_{\max(T_1, t_i)}^{T_2} \left(\frac{t - t_i}{c} + 1\right)^{-p} dt \\
& = (T_2 - T_1)\mu + \sum_{(t_i, m_i)\in \mathcal H}K e^{\alpha(m_i - M_0)}\frac{c}{p-1}\left(\left(\frac{\max(t_i, T_1) - t_i}{c} + 1\right)^{1-p} - \left(\frac{T_2 - t_i}{c} + 1\right)^{1-p}\right) \\ 
& = \Lambda_0(T_1, T_2) + \sum_{(t_i, m_i) \in \mathcal H} \Lambda_i(T_1, T_2)
\end{aligned}
\end{equation}

The above integral can be considered as the sum of two parts, the number of background events $\Lambda_0(T_1, T_2)$ and the remaining summation which is referred as the sum of the number of triggered events by each event $t_i$, namely $\Lambda_i(T_1, T_2)$. We approximate the integral by linearising the functions $\Lambda_0(T_1, T_2)$ and $\Lambda_i(T_1, T_2)$. This means that the resulting approximate integral is the sum of $| \mathcal H | + 1$ linear functions of the parameters. 

However, we concluded that this approximation alone is not sufficiently accurate. To increase the accuracy of the approximation, for each integral $\Lambda_i(T_1, T_2)$, we further consider a partition of the integration interval $[\max(T_1, t_i), T_2]$ in $B_i$ bins, $t^{(b_i)}_0,....,t^{(b_i)}_{B_i}$ such that $t^{(b_i)}_0 = \max(T_1, t_i)$, $t^{(b_i)}_{B_i} = T_2$ and $t^{(b_i)}_j < t^{(b_i)}_k$ if $j < k$. By doing this, the integral becomes, 

\begin{equation} \label{eq:bins_term}
\Lambda(T_1, T_2) = \Lambda_0(T_1, T_2) + \sum_{(t_i, m_i) \in \mathcal H} \sum_{j = 0}^{B_i - 1} \Lambda_i(t^{(b_i)}_j, t^{(b_i)}_{j+1}) 
\end{equation}

In this way, the integral is decomposed in $\sum_i B_i + 1 > | \mathcal H| + 1$ terms providing a more accurate approximation. We discuss the options for temporal binning in Section \ref{binning}.

Substituting Eqn.\ref{eq:bins_term} into \ref{eq:HP_LL}, the Hawkes process log-likelihood can be written,

\begin{equation} \label{eq:allTerms}
\mathcal L(\boldsymbol \theta | \mathcal H) = - \Lambda_0(T_1, T_2) - \sum_{(t_i, m_i) \in \mathcal H} \sum_{j = 0}^{B_i - 1} \Lambda_i(t^{(b_i)}_j, t^{(b_i)}_{j+1}) +  \sum_{(t_i, m_i) \in \mathcal H} \log \lambda(t_i | \mathcal H_{t_i}) .
\end{equation}

In our approximation we linearise the logarithm of each elements within summations with respect to the posterior mode $\boldsymbol \theta^*$. Other choices led to a non-convergent model \cite{serafini2022approximation}. In this case, the approximate log-likelihood becomes,

\begin{equation}
\overline{\mathcal L}(\boldsymbol \theta | \mathcal H) =  - \exp\{\overline{\log\Lambda}_0(T_1, T_2)\} - \sum_{(t_i, m_i) \in \mathcal H} \sum_{j = 0}^{B_i - 1} \exp\{\overline{\log\Lambda}_i(t^{(b_i)}_j, t^{(b_i)}_{j+1})\} +  \sum_{(t_i, m_i) \in \mathcal H} \overline{\log \lambda}(t_i | \mathcal H_{t_i})
\end{equation}

where for a generic function $f(\boldsymbol \theta)$ with argument $\boldsymbol \theta \in \Theta \subset \mathbb R^m$, the linearised version with respect to a point $\boldsymbol \theta^*$ is given by a truncated Taylor expansion,

\begin{equation}
\overline{f}(\boldsymbol \theta) = f(\boldsymbol \theta^*) + \sum_{k = 1}^m (\theta_k - \theta^*_k) \frac{\partial}{\partial \theta_k}f(\boldsymbol \theta) \Bigg\vert_{\boldsymbol \theta = \boldsymbol \theta^*}
\end{equation}

The other key component is the functions for each of the three incremental components in Eqn.\ref{eq:allTerms} that need to be linearised in this way. These are provided in \texttt{ETAS.inlabru} and will be discussed in Section \ref{funcLin}. 

\subsection{Temporal binning} \label{binning}

 For each event, time binning is used to in Eqn.\ref{eq:bins_term} to improve the accuracy of the integration of the term describing the sum of the number of triggered events. The binning strategy is fundamental because the number of bins determines, up to a certain limit, the accuracy of this component of the approximation. Considering more bins enhances the accuracy of the approximation but increases the computational time because it increases the number of quantities to be approximated. Also, we cannot reduce the approximation error to zero, and the numerical value of the integral in each bin goes to zero increasing the number of bins which can be problematic in terms of numerical stability. We found that for the ETAS model considered here, having around $10$ bins for each observed point is usually enough, and that is best considering higher resolution bins close to the triggering event. In fact, the function

\begin{equation}
g_t(t, t_i, m_i) = K e^{\alpha(m_i - M_0)}\left(\frac{t - t_i}{c} + 1\right)^{-p}\mathbb I(t > t_i)
\end{equation}

varies the most for value of $t$ close to $t_i$ and become almost constant moving away from $t_i$. This means that we need shorter bins close to $t_i$, to capture the variation, and wider bins far from $t_i$ where the rate changes more slowly. 

We choose a binning strategy defined by three parameters $\Delta, \delta > 0$, and $n_{max} \in \mathbb N^+$. The bins relative to the observed point $t_i$ are given by

\begin{equation} \label{eq:binning}
t_i, \, t_i + \Delta, \, t_i + \Delta(1 + \delta),\, t_i + \Delta(1 + \delta)^2,....,t_i + \Delta(1 + \delta)^{n_i}, T_2  
\end{equation}

where, $n_i \leq n_{max}$ is the maximum $n \in \{0,1,2,3,...\}$ such that $t_i + \Delta(1 + \delta)^n < T_2$. The parameter $\Delta$ regulates the length of the first bin, $\delta$ regulates the length ratio between consecutive bins, and the value $n_{max}$ regulates the maximum number of bins.

This strategy presents two advantages. The first is that we have shorter bins close to the point $t_i$ and wider bins as we move away from that point. The second is that the first (or second, or third, or any) bin has the same length for all points. This is useful because the integral in a bin is a function of the bin length and not of the absolute position of the bin. This means that we need to calculate the value of the integral in the first (second, third, or any) bin once time and reuse the same result for all events. This significantly reduces the computational burden. 

\subsection{Functions to be linearized} \label{funcLin}

This section and the next one illustrate what we need to provide to \inlabru to approximate Hawkes process models. This one focuses on the functions to be provided while the next one on how they are combined to obtain the desired approximation. Regarding the functions to be provided, we remark that those are already present in the \texttt{ETAS.inlabru} package, so the user does not have to provide anything apart from the data, the area of interest, and the prior parameters. However, these sections are useful to understand what happens under the curtains and if one wants to extend this approach to more complicated ETAS implementations.   

To build an ETAS model, we need to provide functions for each of the components of the likelihood function (Eqn.{\ref{eq:allTerms}}). The linearisation and the finding of the mode $\boldsymbol \theta^*$ are managed automatically by the \texttt{inlabru} package. We only have to provide the functions to be linearised. Specifically, we need to provide the logarithm of the functions needed to approximate the integral and the logarithm of the conditional intensity. More formally, for our approximation of the ETAS model (i.e. for each term in Eqn {\ref{eq:allTerms}}), \texttt{ETAS.inlabru} provides the functions, 

\begin{equation}
\log \Lambda_0(T_1, T_2) = \log(T_2 - T_1) + \log(\mu)
\end{equation}

\begin{align*}
\log\Lambda_i(t^{(b_i)}_j, t^{(b_i)}_{j+1}) = \log(K) & + \alpha(m_i - M_0) + \log\left(\frac{c}{p-1}\right) 
\\ & + \log\left(\left(\frac{t^{(b_i)}_j - t_i}{c} + 1\right)^{1-p} - \left(\frac{t^{(b_i)}_{j+1} - t_i}{c} + 1\right)^{1-p}\right) 
\end{align*}

and 

\begin{equation}
\log\lambda(t|\mathcal H_t) = \log\left(\mu + \sum_{(t_h, m_h) \in \mathcal H_t} K e^{\alpha(m_h - M_0)}\left(\frac{t - t_h}{c} + 1\right)^{-p}\right)
\end{equation}

For full details see \cite{serafini2022approximation}.

\subsection{Implementation Details: The Poisson Count model trick}

Our implementation in \inlabru works by combining three INLA Poisson models on different datasets. The use of the INLA Poisson model here is related to computational efficiency purposes; it does not have any specific statistical meaning. Specifically, we leverage the internal log-likelihood used for Poisson models in R-INLA (and \inlabru) to obtain the approximate Hawkes process log-likelihood as part of a computational trick. 

More formally, INLA has the special feature of allowing the user to work with Poisson counts models with exposures equal to zero (which should be improper). A generic Poisson model for counts $c_i, i = 1,...,n$ observed at locations $\mathbf t_i, i = 1,...,n$ with exposure $E_1,...,E_n$ with log-intensity $\log\lambda_P(\mathbf t) = f(\mathbf t, \boldsymbol \theta)$, in \inlabru has log-likelihood given by:

\begin{equation}
    \mathcal L_P(\boldsymbol \theta) \propto -\sum_{i = 1}^n \exp\{\overline{f}(\mathbf t_i, \boldsymbol \theta, \boldsymbol \theta^*)\}*E_i + \sum_{i=1}^n \overline{f}(\mathbf t_i, \boldsymbol \theta, \boldsymbol \theta^*)*c_i
    \label{eq:23_poissonloglik}
\end{equation}

Each Hawkes process log-likelihood component (Eqn. \ref{eq:allTerms}) is approximated using one surrogate Poisson model with log-likelihood given by Eqn. \ref{eq:23_poissonloglik} and an appropriate choice of counts and exposures data. Table \ref{tab:2_loglikapprox} reports the approximation for each log-likelihood component with details on the surrogate Poisson model used to represent it. For example, the first part (integrated background rate) is represented by a Poisson model with log-intensity $\log\Lambda_0(\mathcal X)$, this will be automatically linearised by \inlabru. Given that, the integrated background rate is just a scalar and not a summation, and therefore we only need one observation to represent it assuming counts equal 0 and exposures equal 1. Table \ref{tab:2_loglikapprox} shows that to represent a Hawkes process model having observed $n$ events, we need $1 + \sum_h (B_h) + n$ events with $B_h$ number of bins in the approximation of the expected number of induced events by observation $h$.

Furthermore, Table \ref{tab:2_loglikapprox} lists the components needed to approximate the ETAS log-likelihood which will be internally considered as surrogate Poisson log-intensities by \inlabru. More specifically, we only need to create the datasets with counts $c_i$, exposures $e_i$, and the information on the events $\mathbf x_i$ representing the different log-likelihood components; and, to provide the functions $\log\Lambda_0(\mathcal X), \log\Lambda_h(b_{i,h}),$ and, $\log\lambda(\mathbf t)$. The linearisation is automatically performed by \inlabru as well as the retrieving of the parameters' posterior distribution.

\begin{table}
  \centering
  \scalebox{0.8}{
  \begin{tabular}{llllll}
    \toprule
    %\cmidrule(r){1}
    Name     & Objective   & Approximation & Surrogate $\log\lambda_P$ & Number of data points & Counts and Exposures \\
    \midrule
    Part I & $\Lambda_0(\mathcal X)$ & $\exp\overline{\log\Lambda}_0(\mathcal X)$ & $\log\Lambda_0(\mathcal X)$ & 1 & $c_i = 0$, $e_i = 1$       \\ \\
    Part II & $\sum_{h=1}^n\sum_{i = 1}^{B_h} \Lambda_h(b_{i, h})$ & $\sum_{h=1}^n\sum_{i = 1}^{B_h} \exp\overline{\log\Lambda}_h(b_{i,h})$ & $\log\Lambda_h(b_{i,h})$ & $\sum_h B_h$ & $c_i = 0$, $e_i = 1$       \\ \\
    Part III & $\sum_{h=1}^n\log\lambda(\mathbf x_h)$ & $\sum_{h=1}^n \exp\overline{\log\lambda}(\mathbf x_h)$ & $\log\lambda(\mathbf x)$ & $n$ & $c_i = 1$, $e_i = 0$       \\ \\
    \bottomrule
  \end{tabular} }
  \caption{Hawkes process log-likelihood components approximation}
  \label{tab:2_loglikapprox}
\end{table}

More detail on how to build the functions in the \texttt{ETAS.inlabru} package can be found at \url{https://github.com/Serra314/Hawkes_process_tutorials/tree/main/how_to_build_Hawkes}.

\subsection{Prior specification} \label{priors}

We have to set the priors for the parameters. The INLA method is designed for Latent Gaussian models, which means that all the unobservable parameters have to be Gaussian. This seems in contrast with the positivity constraint of the ETAS parameters $\mu, K, \alpha, c, p$, but we have a solution. 

Our idea is to use an internal scale where the parameters have a Gaussian distribution and to transform them before using them in the log-likelihood components calculations. We refer to the internal scale as INLA scale, and to the parameters in the INLA scale as $\boldsymbol\theta$. In practice, all parameters have a standard Gaussian prior in the INLA scale and they are transformed to be distributed according to a target distribution in the ETAS scale. Specifically, assuming that $\theta$ has a standard Gaussian distribution with cumulative distribution function (CDF) $\Phi(\theta)$, and calling $F^{-1}_Y$ the inverse of the CDF of the target distribution for the parameter, we can switch between the Gaussian and the target distributions using,

\begin{equation}
\eta(\theta) = F^{-1}_Y(\Phi(\theta))    
\end{equation}

where $\eta(\theta)$ has a distribution with CDF $F_Y(\cdot)$.

The \texttt{ETAS.inlabru} R-package uses the following default priors in the ETAS scale, 

\begin{align}
    \mu_b & \sim \text{Gamma}(\text{shape} = 0.5, \text{rate} = 0.5) \nonumber\\
    K_b & \sim \text{LogNormal}(\text{mean}(\log(K))=-1, \text{sd}(\log(K))=0.5) \nonumber\\
    \alpha_b & \sim \text{Unif}(\alpha_{min} = 0, \alpha_{max} = 10) 
    \label{eq:31_lnpriors} \\
    c_b & \sim \text{Unif}(c_{min} = 0, c_{max}=1) \nonumber\\
    p_b & \sim \text{Unif}(p_{min}=1, p_{max}=2) \nonumber 
\end{align}
however, they can be changed to different distributions that better describe the available prior information. 

The package \inlabru provides a function to easily implement such transformation. The function is called \texttt{bru\_forward\_transformation} and takes in input the quantile function of the target distribution and its parameters. Below we report three examples of transformations such that the parameters in the ETAS scale have a Gamma, Uniform, or Log-Gaussian distribution. We show the empirical density obtained by transforming the same sample of values from a standard Gaussian distribution. 

The prior for $\mu$ is the one that will most commonly need to be modified as it changes with the size of the domain being modelled. We choose  $\text{Gamma}(\text{shape}=a_\mu , \text{rate}=b_\mu))$ prior for $\mu$. The mean of the distribution is given by $a_\mu / b_\mu =1$ event/day, the variance is $a_\mu / b^2_\mu = 2$ and the skewness $2/\sqrt\alpha = 2.5$. One strategy for setting these parameters is to estimate an upper limit on the rate by dividing the duration of the catalogue total number of events; this is likely an overestimate as it combines the triggered and background events. One might choose to pick a mean rate that is ~half of this which defines the ratio of $a_\mu$ and $b_\mu$. There is then some trade-off in the variance and skewness parameters.

\begin{figure}[h!]
\begin{center}
\includegraphics[width=15cm]{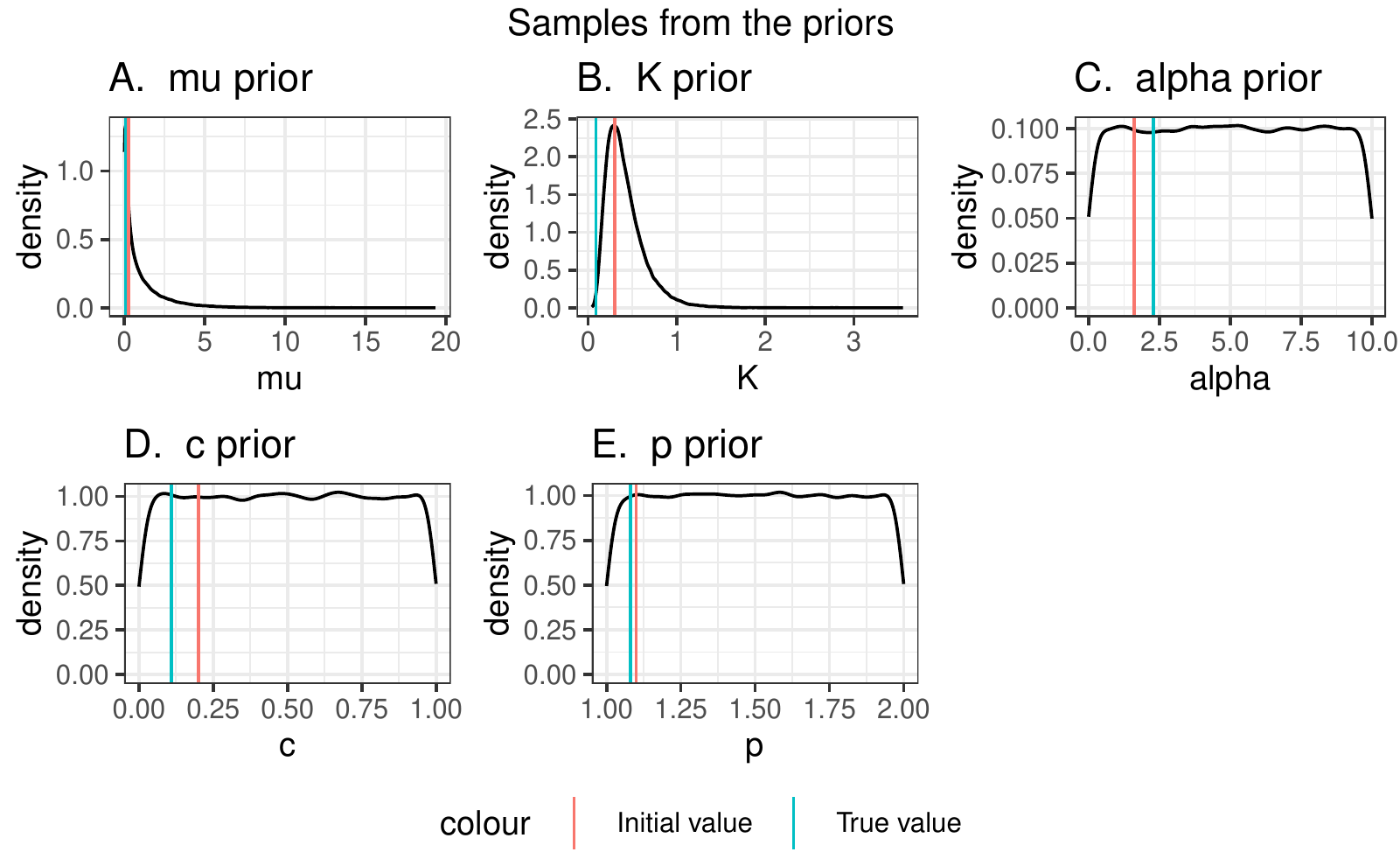}
\end{center}
\caption{Plot showing samples from the priors on the ETAS scale that we use throughout this paper. They are intentionally broad. The red line shows the initial value used for the majority of the analyses in this paper and the green line shows the true value. }\label{fig:priors}
\end{figure}

Samples drawn from the priors used in this paper are shown in Figure \ref{fig:priors}, including lines showing the initial and true values that will be used through the majority of the results. The sensitivity to the choice of initial values is the first part of the results section. Please note how broad these priors are as this is helpful to see how much more informative the posteriors we generate are from these initial distributions.

\begin{figure}[h!]
\begin{center}
\includegraphics[width=15cm]{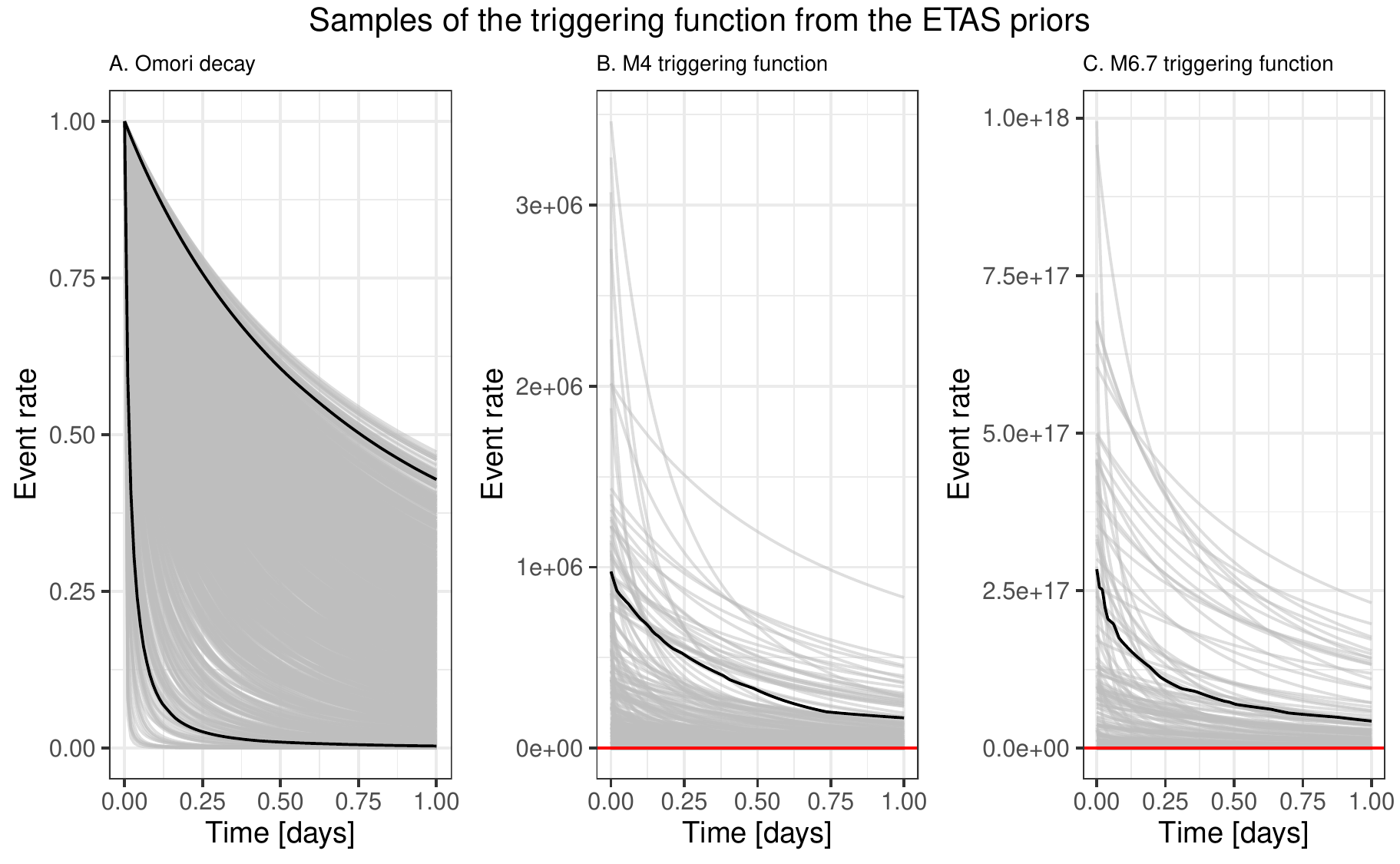}
\end{center}
\caption{Plot samples of the triggering functions drawn from the priors. }\label{fig:priors_trig}
\end{figure}

The ETAS parameters themselves are not easy to interpret given that it is their combination in the Omori decay and triggering functions that we are most interested in. We draw 1000 samples from the priors to generate samples of the Omori decay, the triggering function for an M4 event and the triggering function of an M6.7 event (Fig.\ref{fig:priors_trig}). We see that these priors produce a wide range of behaviour, including unrealistically large productivity compared to real earthquake process. This information is useful for comparison with the triggering functions derived from sampling posteriors later in the paper.

\subsection{Fitting the model}

The function \texttt{Temporal.ETAS.fit(list.input)} performs the ETAS inversion. The \texttt{list.input} object is a structured list containing the raw catalogue, the catalogue formatted for \inlabru, definition of the model domain, an initial set of trail parameters on the ETAS scale, the link functions used to transform from the internal scale to the ETAS scale, parameters to set each of the priors, parameters to generate the time binning, and a series of runtime parameters that control the behaviour of \inlabru. There is a complete description of the parameters in Table \ref{tab:list.input} which cross-references to the section of the paper that describes their role.

In the results section, we vary the catalog, start times, and the initial set of trial parameters. To achieve this, we create a default \texttt{list.input} object and then modify these inputs by hand - the notebooks provided demonstrate how to do this.

\begin{table}
  \centering
  \scalebox{0.8}{
  \begin{tabular}{p{0.22\linewidth}|p{0.15\linewidth}|p{0.63\linewidth}}
    \toprule
    %\cmidrule(r){1}
    Parameter and Type & Default Value & Further information \\
    \midrule
    \textbf{Data} & & Catalogue of event times and magnitudes\\
    \, catalogue  $[t,M(,...)]$ & & The input catalog as it is provided with at least a set of times and magnitudes \\
    \, catalogue.bru  list([ts, magnitude, idx.p])& & The input catalog in the format needed for inlabru. For each event we have a [time, magnitude, id] \\ \midrule
    \textbf{Domain Definition}  & & Time domain varied in Sections \ref{representative} and \ref{historic}\\ 
    \, time.int  & & The provided start and end date in string format \\
    \, T12  double [T1, T2] & & The start and end date as number of days from the provided starting date \\
    \, lat.int  double  & [-90,90] & Min and max latitude bounds for filtering the catalogue \\
    \, lon.int  double  & [-180,180] & Min and max longitude bounds for filtering the catalogue \\
    \, M0  double & 2.5 & Minimum magnitude for the model domain \\
    \midrule
    \textbf{Initial trial paras}  & & Varied in Section \ref{varyInit} \\
    \, mu.init  double & 0.3 & Initial guess for the background rate, $\mu$ \\
    \, K.init  double  & 0.1 & Initial guess for the, $K$\\
    \, alpha.init  double & 1 & Initial guess for, $\alpha$ \\
    \, c.init  double & 0.2 & Initial guess for, $c$ \\
    \, p.init  double & 1.1  & Initial guess for, $p$ \\
    \midrule
    \textbf{Link functions}  & & A list of functions used to transform the parameters from the internal scale to the ETAS scale\\
    \midrule
    \textbf{Priors}  & & See Section \ref{priors} for definition\\
    \, a\_mu  double & 0.5 & Gamma distribution shape parameter \\
    \, b\_mu  double & 0.5 & Gamma distribution rate parameter \\
    \, a\_K  double & -1 &  log-Normal distribution mean\\
    \, b\_K  double & 0.5 &  log-Normal distribution standard deviation\\
    \, a\_alpha  double & 0 & min of a uniform distribution \\
    \, b\_alpha  double & 10 & max of a uniform distribution \\
    \, a\_c  double & 0 & min of a uniform distribution \\
    \, b\_c  double & 1 & max of a uniform distribution \\
    \, a\_p  double & 1 & min of a uniform distribution \\
    \, b\_p  double & 2 & max of a uniform distribution \\
    \midrule
    \textbf{Time binning paras} & & See Section \ref{binning} \\
    \, Nmax  int & 8 & value of the parameter $n_{max}$ in Eqn.\ref{eq:binning} \\  
    \, coef.t  double & 1.0 & value of the parameter $\delta$ in Eqn. \ref{eq:binning} \\  
    \, delta.t  double & 0.1 & value of the parameter $\Delta$ in Eqn. \ref{eq:binning} \\  
    \midrule
    \textbf{bru.opt.list}  & & See bru documentation \\
    \, bru.verbose  int & 3 & type of visual output from \inlabru\\  
    \, bru\_max\_iter  int & 100 & maximum number of \inlabru iterations \\  
    \, num.threads  int & 5 &  number of cores used in each \inlabru iteration\\ 
    \, inla.mode  string & 'experimental' & type of approximation used by INLA \\
    \, bru.inital: th.mu, th.K, \\ \,\, th.alpha, th.c, th.p & list[double[5]] & Initial trial parameters on the internal scale. These are calulated using the inverse of the copula transformation functions in \texttt{ETAS.inlabru} \\
    \midrule
    \textbf{Runtime paras}  & & \\
    \, max\_iter  int & 100 & maximum number of iterations for the inlabru algorithm. The number of iterations will be less than this number if the algorithm have converged \\
    \, max\_step  &  NULL & this parameters refers to how far the parameter value can jump from one iteration to another. The greater the value the greater the potential jump. Setting a value different from NULL prevents the \inlabru algorithm to check for convergence and the algorithm will run exactly the number of iterations specified in $max\_iter$. \\ 
    \hline
  \end{tabular} }
  \caption{Description of the model definition contained in list.input. This information will be passed to \texttt{ETAS.inlabru} to start the inversion. Each analysis in the results section is initialised by adjusting this list.}
  \label{tab:list.input}
\end{table}

Once we call \texttt{Temporal.ETAS.fit(list.input)}, the iterative fitting of the model parameters is handled automatically by \inlabru until they converge or $max\_iter$ iterations have occurred. For a comprehensive discussion of the underlying mathematical framework, we refer the reader to \cite{serafini2022approximation}. 

\begin{figure}[h!]
\begin{center}
\includegraphics[width=15cm]{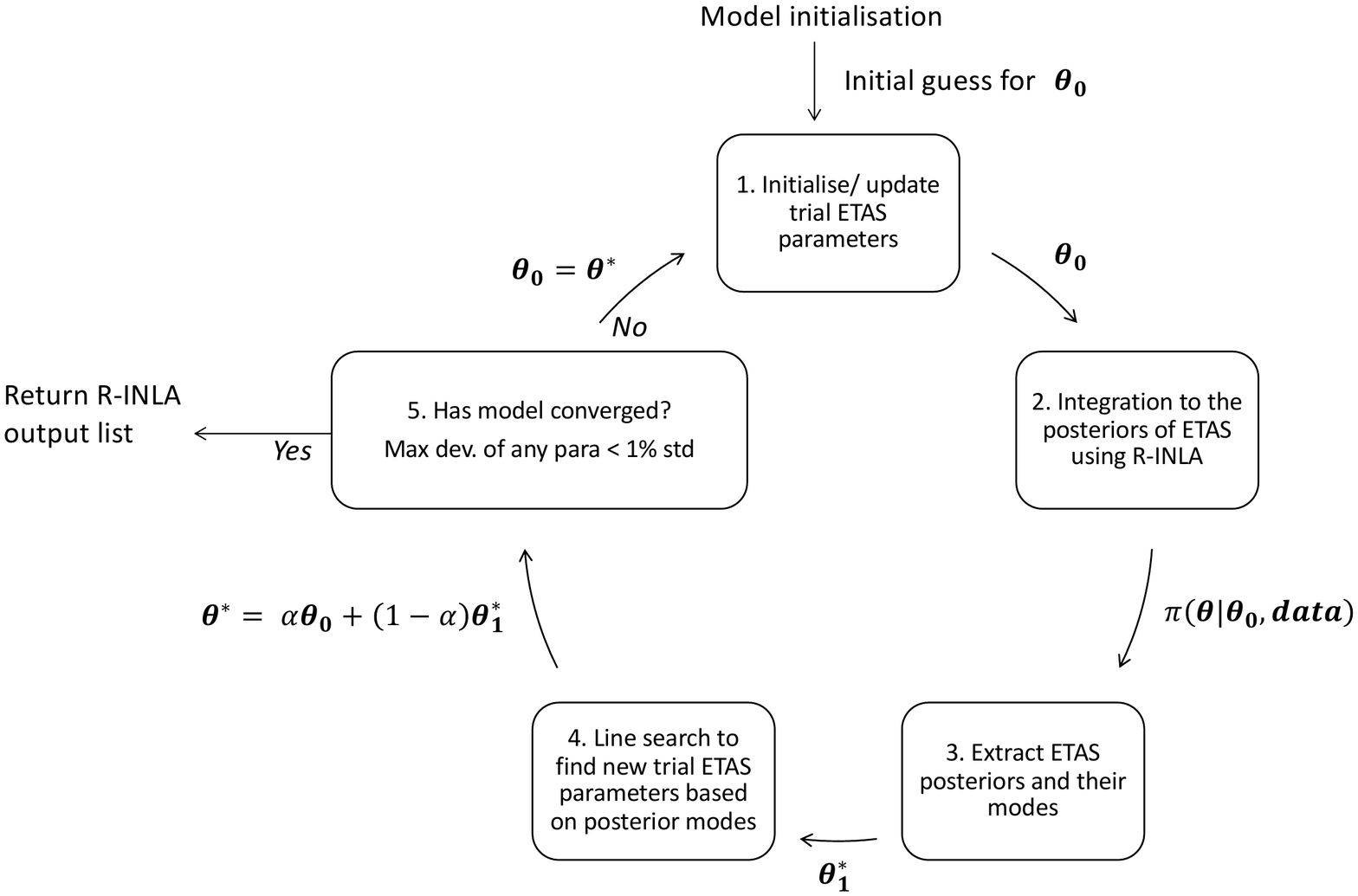}
\end{center}
\caption{Schematic diagram showing the \inlabru workflow which iteratively updates a set of trial ETAS parameters.}\label{fig:1}
\end{figure}

 The iterative process is illustrated in Figure \ref{fig:1} and outline each of the steps below.

\subsubsection{Step 1: Initialise/update trial ETAS parameter set}

We start with a set of trial ETAS parameters $\boldsymbol \theta_0 = (\mu_0, K_0, \alpha_0, p_0, c_0)$ which will be used as the linearisation point for the linear approximation. These initial values should lie within their respective priors. They could be sampled from the priors, but it is possible that a very unrealistic parameter combination might be chosen. These parameters will be updated each loop of the \inlabru algorithm. In general, extreme parametrisations (e.g. parameters smaller than $10^{-5}$ or greater than $20$) should be avoided. Usually, setting all the parameters to 1 (expect $p$ which could be set to 1.1) is a safe choice. Another approach could be to use the maximum likelihood estimate.  

\subsubsection{Step 2. Integrated Nested Laplace Approximation}

\texttt{ETAS.inlabru} contains the ETAS functions that will be internally linearised (see Section \ref{funcLin}) about an arbitrary point and then integrated. The nested integration is performed by R-INLA, but this is managed by \inlabru so we never need to call it directly. The R-INLA output returns a comprehensive output, including the joint posteriors (LINK TO R-INLA output doc).

\subsubsection{Step 3. Extract the ETAS posteriors and their modes}

From the R-INLA output, we extract the modes of the approximated posteriors $\boldsymbol \theta_1^*$. In early iterations, this point is usually far away from the true mode posterior, this depends on the point $\boldsymbol\theta_0$ used as starting point. The approximate posterior mode tends to the true one as the iterations run. 

\subsubsection{Step 4. Line search to update modal parameters}

At this point we have the initial set of trial ETAS parameters $\boldsymbol \theta_0$ that were used as the linearisation point, and the posterior modes derived from R-INLA $\boldsymbol\theta^*_1$. The value of the linearisation point is updated to $\boldsymbol \theta^* = \alpha \boldsymbol\theta_0 + (1-\alpha)\boldsymbol\theta_1^*$, where the scaling $\alpha$ is determined by the line search method described here \url{https://inlabru-org.github.io/inlabru/articles/method.html}.

\subsubsection{Step 5. Evaluation of convergence}

Convergence is evaluated by comparing $\boldsymbol \theta^*$ and $\boldsymbol \theta_0$. By default, convergence is established when there is a difference between each parameter pair is less than a 1\% of the parameter standard deviation. The value $1\%$ can be modified by the user. If convergence has not been achieved and the maximum number of iterations have not occurred, we set $\boldsymbol\theta_0 = \boldsymbol\theta^*$ and return to step 1 using the new linearisation point as the set of trial parameters.

\subsection{Generation of synthetic catalogues}

The final component of this paper is the production of synthetic catalogues to be analysed. The synthetics are constructed leveraging the branching structure of the ETAS model. Specifically, for temporal models, background events are selected randomly in the provided time window with a rate equal to $\mu$. Then, the offsprings of each background event are sampled from an inhomogeneous Poisson process with intensity given by the triggering function. This process is repeated until no offsprings are generated in the time frame chosen for the simulation.

Using \texttt{ETAS.inlabru}, we generate catalogues with a duration of 1000 days with a background rate of $\mu=0.1$ events per day and ETAS triggering parameters of $c=0.11$, $p=1.08$, $\alpha=2.29$ and $K=0.089$. We take a $b$-value of 1 for the Gutenberg-Richter magnitude distribution. The lower magnitude threshold $M_0=2.5$ which is motivated by catalogues such as those in the Appenines of Italy or for Ridgecrest in California.

We also use two different scenarios, a seeded version of these catalogues where we impose an extra M6.7 event on day 500, and an unseeded catalogue where the events are purely generated by the ETAS model. This leads to catalogues which are relatively active in the former case and relatively quiet in the latter case.  Using these scenarios, we can generate different stochastic samples of events to produce a range of catalogues consistent with these parameterisations.

The R Markdown notebook in Supplemental material allow the reader to see how we have implemented these catalogues for the range of models investigated in the results.

\section{Results}

We present the performance of the \inlabru ETAS inversion across a range of synthetic case studies motivated by various challenges of analysing real earthquake catalogues. We are interested in the accuracy and precision of the inversion compared to the original ETAS parameterisation, understanding sources of systematic bias derived from differences in the catalogues being modelled, and the computational efficiency of the method.

All of the analyses start with one or more catalogues of 1000 days in length, generated using a constant background rate of $\mu=0.1$ events/day above a constant magnitude threshold of $M_0=2.5$, and true ETAS parameters listed on the top row of Table \ref{tab:initialConditions}. We choose to use this minimum magnitude as it is equivalent to the real case study examples of California and l'Aquila, Abruzzo, Italy we will return to this in the discussion section.

A consequence of choosing the 1000 day window is that there will not be a fixed number of events when comparing different samples, as some samples contain large events whilst others are relatively quiet. We make this choice because we believe that it represents the closest analogy to the data challenge faced by practitioners. However, we will be explicit in exploring the implications of this choice.

Within the sequences, there are three different timescales or frequencies that inter-relate. The duration of the synthetic catalogues, the background rate and the rate at which aftershocks decay (e.g. \cite{touati2009origin}). A short catalogue would be one which only samples several background events or perhaps a single mainshock aftershock sequence, or less. A long catalogue would contain periods dominated by small background events and also separate periods containing relatively isolated mainshock aftershock sequences. Clearly there is scope for a whole range of behaviour in between. Given that the accuracy of the ETAS inversion is conditional on the catalogue, we should therefore expect factors such as the catalogue duration, rate of background events and the presence of large events to influence the ability of the algorithms to find accurate solutions.

\subsection{Impact of varying the initial trial ETAS parameter set, $\theta_0$} \label{varyInit}

\begin{figure}[h!]
\begin{center}
\includegraphics[width=15cm]{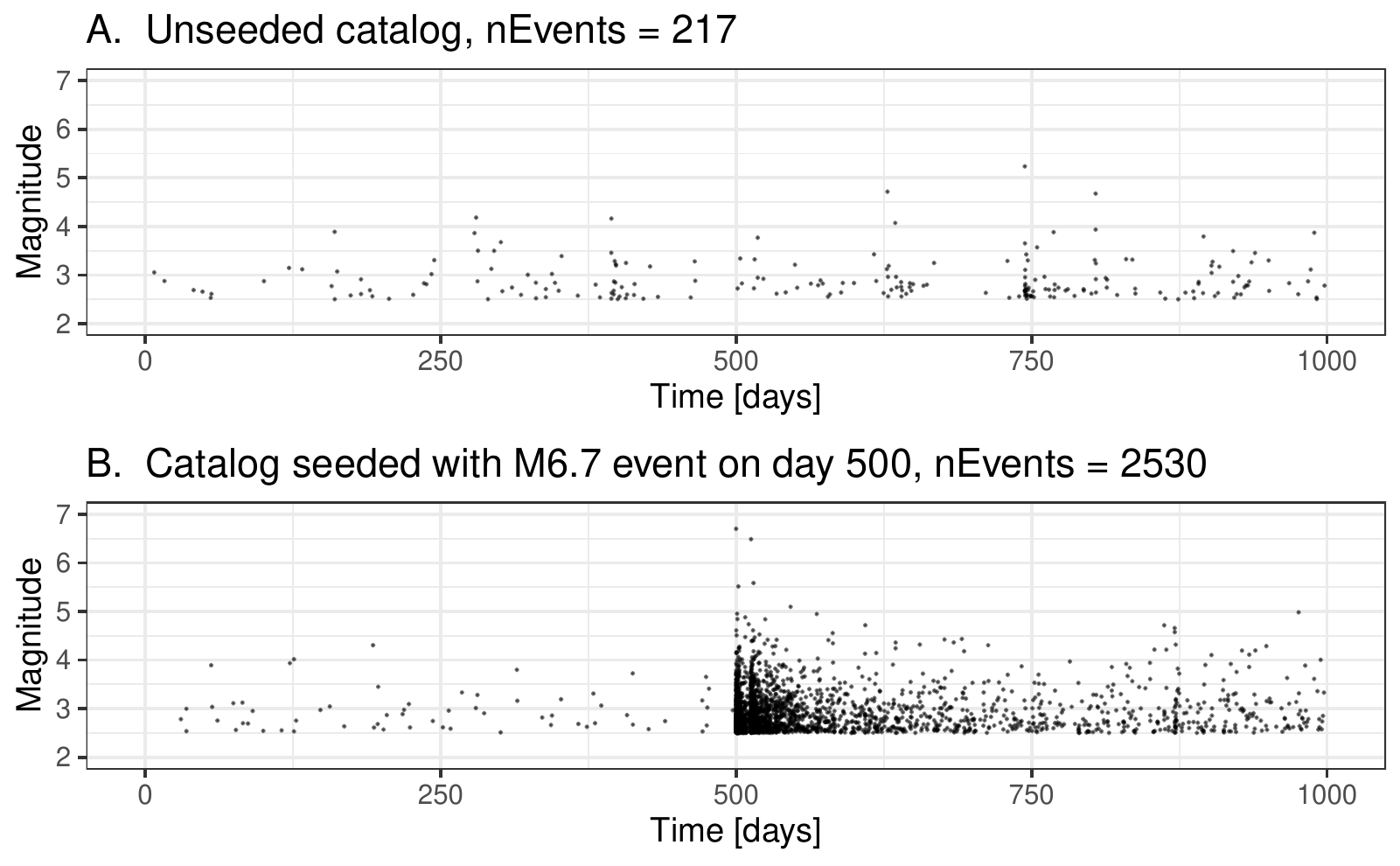}
\end{center}
\caption{The two catalogues we use when varying the starting point for the ETAS parameters.}\label{fig:startingPointCats}
\end{figure}

Where algorithms require an initial set of trial starting parameters, it is important to test whether the results are robust irrespective of the choice of starting conditions. We explore the influence of the initial conditions by generating two catalogue (See Fig.\ref{fig:startingPointCats}) using the ETAS model with the true parameters given on the top row of Table \ref{tab:initialConditions}. They are both 1000 days long and the second catalogue has a M6.7 event seeded on day 500 to produce a more active sequence (Fig.\ref{fig:startingPointCats}(B)). We then invert each catalogue using the different sets of trial ETAS parameters also given in Table \ref{tab:initialConditions}; the third set of initial parameters includes the true solution.

\begin{table}[]
    \centering
    \begin{tabular}{c || c | c | c | c | c }
        Parameter set & $\mu$ & $K$ & $\alpha$ & $c$ & $p$ \\
        \hline
        True parameters & 0.1 & 0.089 & 2.29 & 0.11 & 1.08 \\
        Trial parameter set 1 & 0.05 & 0.01 & 1. & 0.05 & 1.01\\
        Trial parameter set 2 &  5.0 & 1. & 5. & 0.3 & 1.5 \\
        Trial parameter set 3 & 0.1 & 0.089 & 2.29 & 0.11 & 1.08 \\
        Trial parameter set 4 & 0.3 & 0.1 & 1. & 0.2 & 1.01 
    \end{tabular}
    \caption{Table showing the true ETAS parameters and the 4 sets of different initial conditions used in analysing the catalogues in Figure \ref{fig:startingPointCats} to produce the ETAS posteriors in Figure \ref{fig:startingPointPosteriors}.}
    \label{tab:initialConditions}
\end{table}

\begin{figure}[h!]
\begin{center}
\includegraphics[width=15cm]{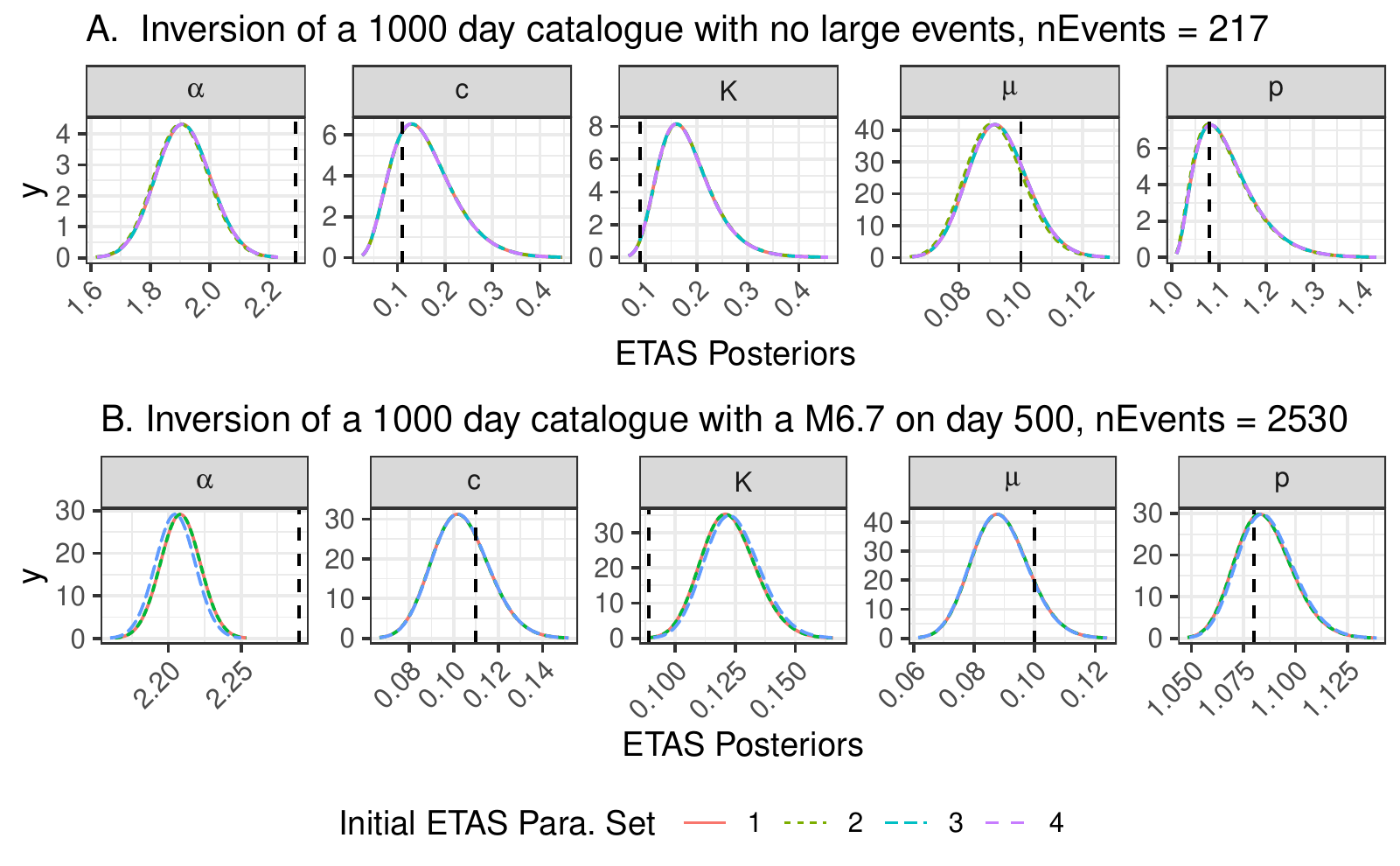}
\end{center}
\caption{The posteriors for the inversion of the catalogues in Fig.\ref{fig:startingPointCats} given 4 different starting points. The vertical black line shows the true values used when generating the synthetic catalogues. Note the very different scales on the x-axes.}\label{fig:startingPointPosteriors}
\end{figure}

The first catalogue is relatively quiet and has only 217 events (Fig. \ref{fig:startingPointCats}(A)). All four sets of initial trial parameters find the same posteriors (Fig.\ref{fig:startingPointPosteriors}(A)). \inlabru provides a good estimate of the background rate $\mu$ for this catalogue. The other posteriors are the parameters that govern the rate of self-exciting triggering. These posteriors are all skew and some of the posteriors are strongly influenced by the their priors, for example the posterior for $p$ spans the entire range of its prior (compare Fig.\ref{fig:priors} for the priors and Fig 4A for the posteriors). The posteriors for the triggering parameters are relatively broad  because the data is not sufficient to produce a narrow likelihood function.

In the second catalogue we have seeded a M6.7 event on day 500 (Fig. \ref{fig:startingPointCats}(B)). This catalogue has a well defined aftershock sequence and therefore contains significantly more events; 2530 events in total. Again, all four sets of initial trial parameters find the same posteriors (Fig.\ref{fig:startingPointPosteriors}(B)). The posterior for the background rate, $\mu$, remains well resolved and there is no reduction in its standard deviation; this indicates that both catalogues have sufficient information to resolve the background rate, even though they are dominated by aftershocks. All of the posteriors for the triggering parameters are significantly narrower than for the first unseeded catalogue. This is down to two factors, firstly there are many more events in the seeded catalogue, and secondly the well resolved aftershock sequence makes it much easier to constrain the triggering parameters.

All of these models find similar posteriors irrespective of the initial trial ETAS parameters set. It is important that the priors are set broad enough to allow the potential for the posteriors to resolve the true value. This is particularly evident for the quieter model where the posteriors on the triggering parameter's rely on more information from the priors.

In real catalogues, the prior for the background rate needs to be set with care because, when considering a purely temporal model, it will vary depending upon the spatial extent being considered; i.e. the background rate of a temporal model when considering a global dataset would be significantly higher than just California. Further, changing the lower magnitude limit significantly changes the total number of background events.

\begin{figure}[h!]
\begin{center}
\includegraphics[width=15cm]{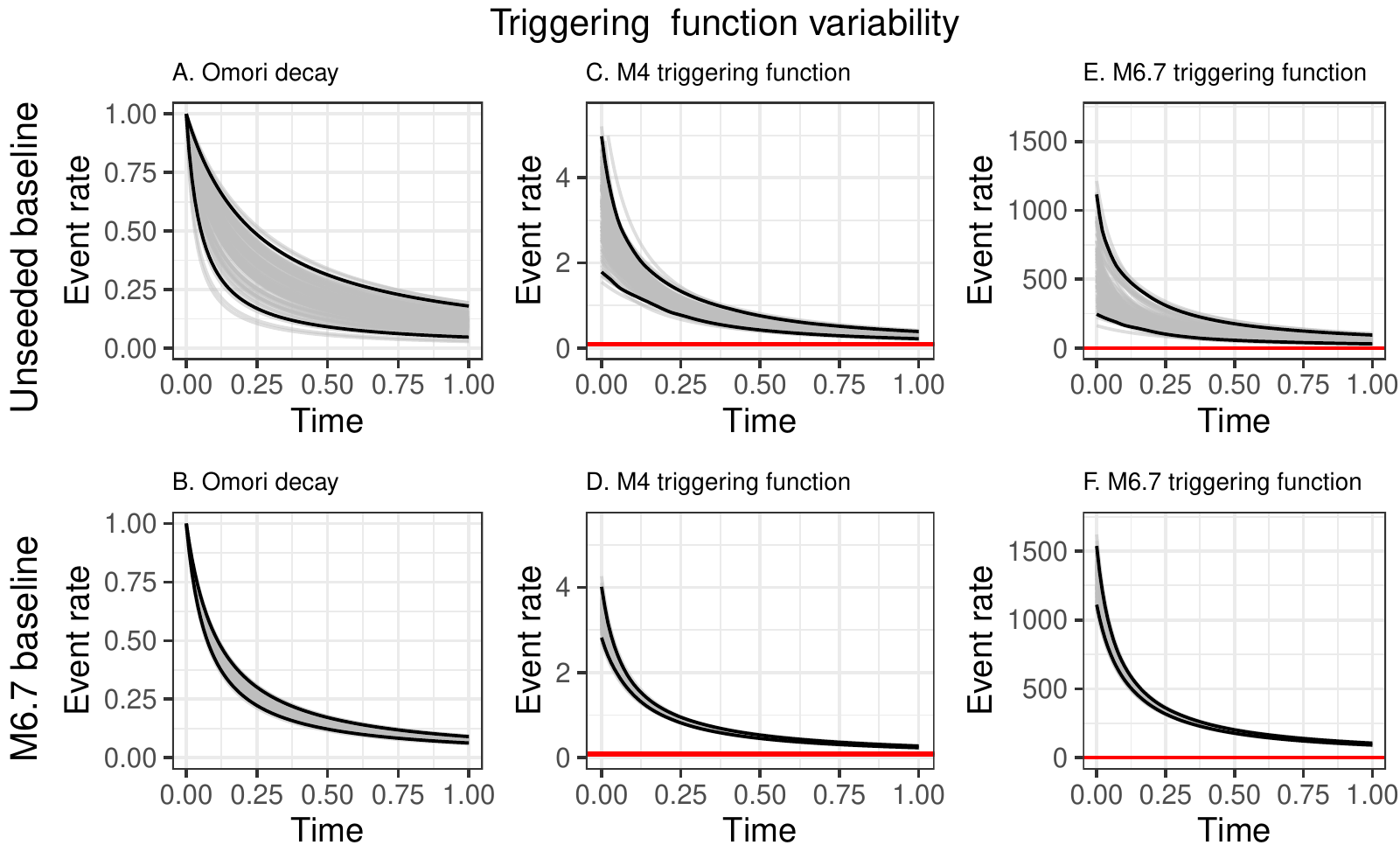}
\end{center}
\caption{Propagation of ETAS parameter uncertainty on the triggering functions. We take 100 samples of the ETAS posteriors for the 1000 day quiescent baseline (top row) and the 1000 day catalogue with an M6.7 on day 500 (bottom row) and use these samples to explore variability in the Omori decay (left hand column), the time-triggering function following an M4 event, and the  time-triggering function following an M6.7 event.}\label{fig:triggering_baseline}
\end{figure}

It is difficult to interpret the ETAS posteriors directly, so we explore the triggering function by sampling the parameter posteriors 100 times, calculating the triggering functions for these posterior samples, and plotting the ensemble of triggering functions (Fig.\ref{fig:triggering_baseline}). The first column shows the Omori function which is the temporal decay of the triggering function, but without the magnitude dependent term - and is therefore also independent of the ETAS $\alpha$ parameter. The larger uncertainty in the posteriors for the unseeded quieter catalogue (Fig.\ref{fig:startingPointCats}(A)) propagate through to much larger variability in the Omori decay (Fig.\ref{fig:triggering_baseline}(A)) than when the sequence is seeded with the large event in the sequence (Fig.\ref{fig:triggering_baseline}(B)). It is reassuring to see that the Omori decay from the sequence seeded with the M6.7 event lies within the confidence intervals of that derived from the quieter unseeded sequence. This implies that the prior is sufficiently broad to capture these extremes in catalogue type. 

When incorporating the magnitude dependence, any bias or uncertainty in $\alpha$ becomes important. The figures show the triggering functions for magnitude 4.0 and 6.7 events over 24 hours. Whilst the triggering functions for the M4 events are nested similar to the Omori sequences, the triggering functions for the M6.7 event are systematically different. The posteriors from the training catalogue seeded with an M6.7 event result an initial event rate ~50\% higher and the two distributions barely overlap. 

We conclude that the choice of training data could have a significant effect on the forecasts of seismicity rate after large events. In the next section we explore the robustness of these results to stochastic uncertainty in the training catalogues.

\subsection{Impact of stochastic variability}

We extend the analysis of the previous section to explore the impact of stochastic variability. We produce 10 synthetic catalogues for both the unseeded and M6.7 seeded catalogues and compare the posterior distributions of each parameter.

\begin{figure}[h!]
\begin{center}
\includegraphics[width=15cm]{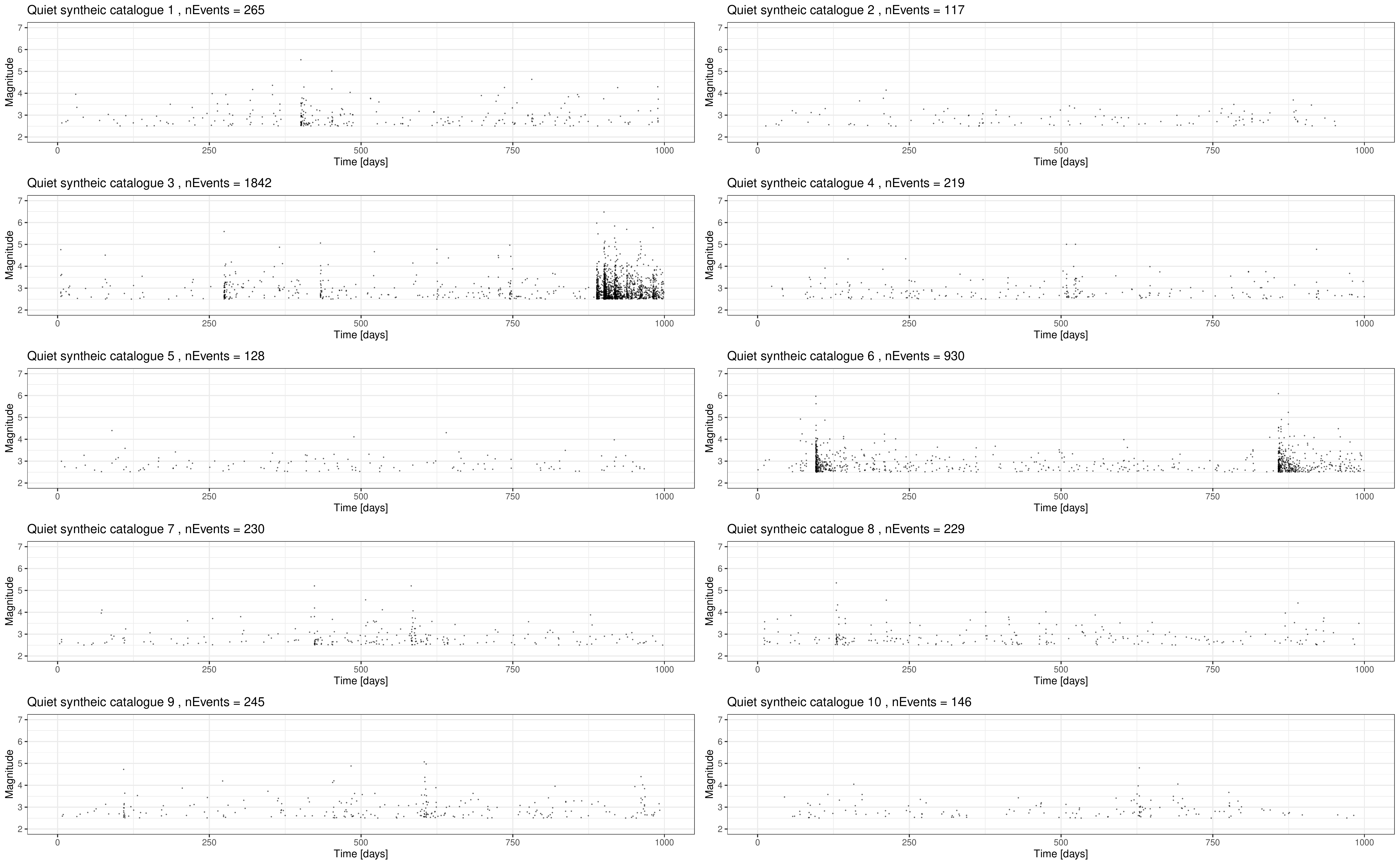}
\end{center}
\caption{10 synthetic catalogues based on the baseline model of 1000 days with background events but no seeded large event. All parameters are the same between the runs and these just capture the stochastic uncertainty. These are all inverted using \inlabru and the family of posteriors is presented in Fig.\ref{fig:incomplete_cat}.}\label{fig:tenQuietCats}
\end{figure}

In the family of catalogues where we did not seed large event (Fig.\ref{fig:tenQuietCats}), we see posteriors of the background rate, $\mu$, that are distributed about the true background rate (Fig.\ref{fig:post_stochastic}(A)) and capture it well. In contrast, we mostly see very large uncertainty in the posteriors for the triggering parameters. However, the true values generally still lie within these posteriors. The very broad posteriors correspond to catalogues that had very few triggered sequences in them. Such broad posteriors illustrate how the Bayesian approach enables us to see where the data did not have sufficient power to narrow the priors significantly; this is useful in evaluating the robustness of a fit. Moreover, the large posterior uncertainty on the parameters would propagate through to large uncertainty in the triggering function if used within a forecast with a rigorous quantification of uncertainty. Synthetic catalogues 3 and 6 (Fig.\ref{fig:tenQuietCats}) contain the largest number of events (1842 and 930 events respectively) as a result of the events triggered by a large random event; these cases have correspondingly tighter and more accurate posteriors for the triggering parameters in (Fig.\ref{fig:post_stochastic}(A)). Similarly, catalogues 1 and 9 have the next highest number of events (265 and 245 events respectively), and these also have the next most informative posteriors. Catalogues 2 and 5 have the fewest events (117 and 128 respectively) and produce posteriors that are significantly informed by the priors, as can be seen the the range of values being explored.

\begin{figure}[h!]
\begin{center}
\includegraphics[width=15cm]{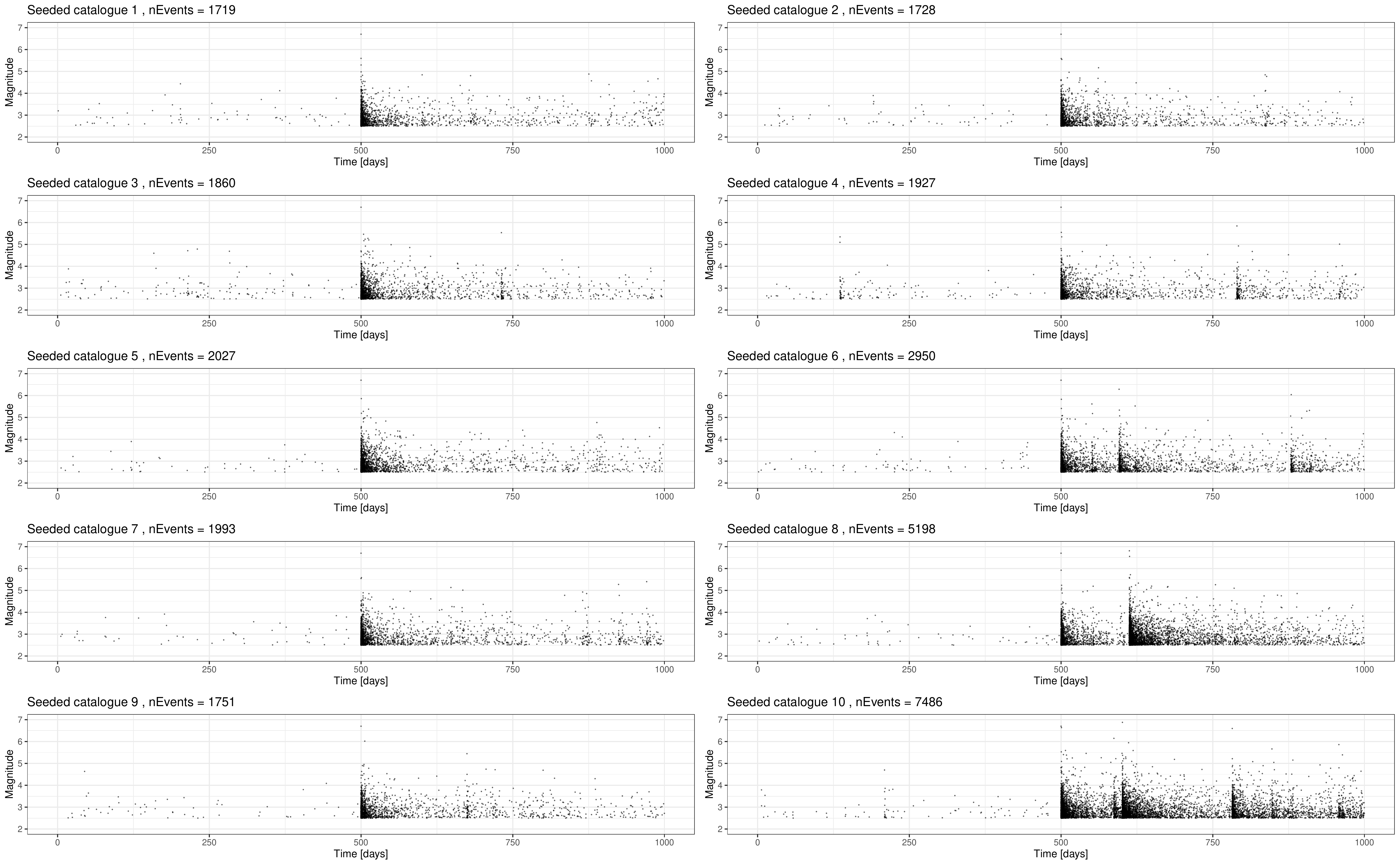}
\end{center}
\caption{10 synthetic catalogues based on the baseline model of 1000 days with background events and a M6.7 event on day 500. All parameters are the same between the runs and these just capture the stochastic uncertainty. These are all inverted using \inlabru and the family of posteriors is presented in Fig.\ref{fig:incomplete_cat}.}\label{fig:10M6p7Cats}
\end{figure}

\begin{figure}[h!]
\begin{center}
\includegraphics[width=15cm]{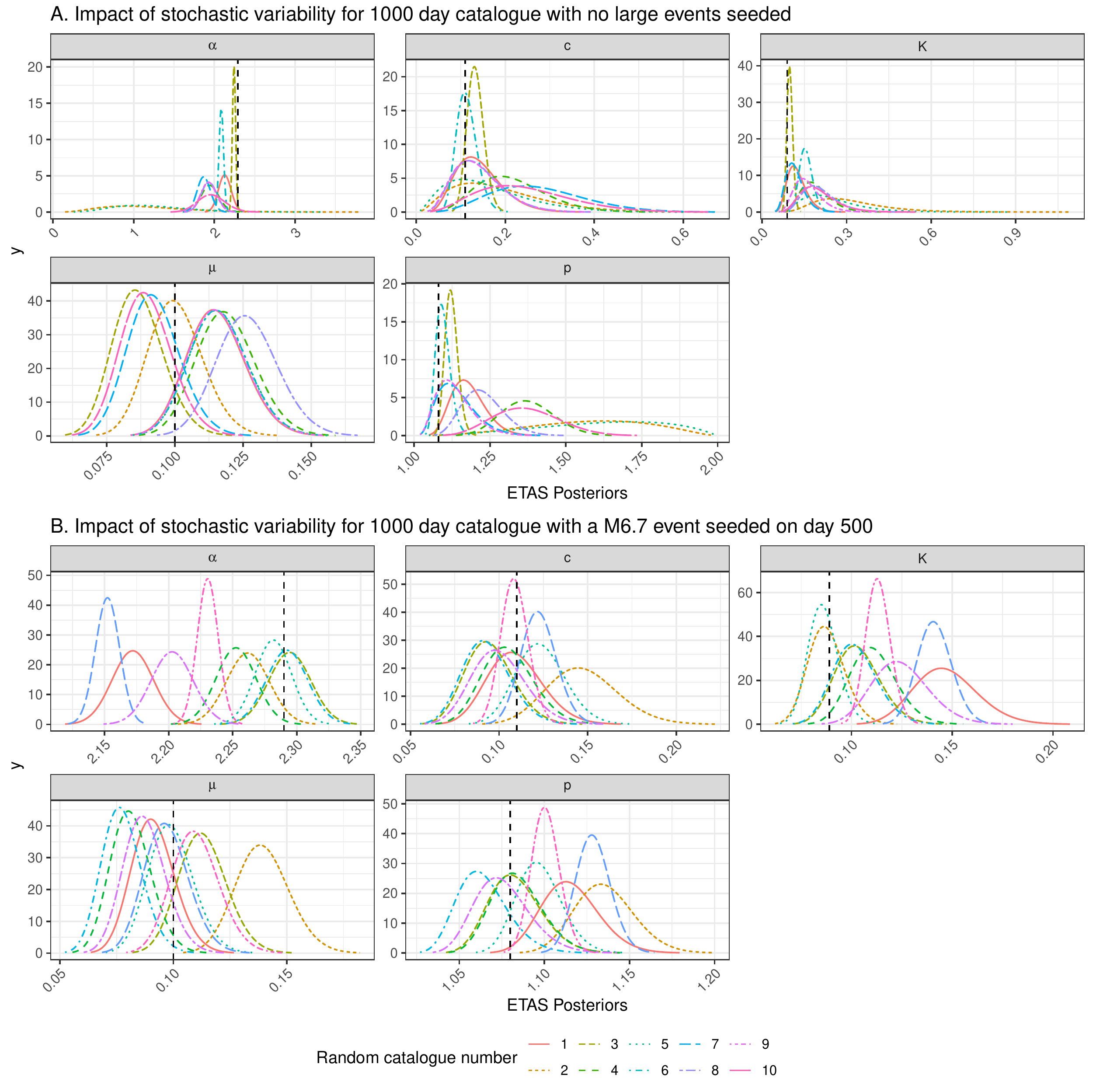}
\end{center}
\caption{Posteriors that explore the impact of stochastic variability on the inverted ETAS parameters using \inlabru. These are based on the relatively quiet catalogues in Figs.\ref{fig:tenQuietCats} and those with a large event seeded on day 500 in Fig.\ref{fig:10M6p7Cats}. The catalogue numbers can be cross-referenced between the figures.}\label{fig:post_stochastic}
\end{figure}

Considering the 10 seeded catalogues (Fig.\ref{fig:10M6p7Cats}), we see a complementary story in the posteriors (Fig.\ref{fig:post_stochastic}(B)). Again, the posteriors for the background rate are distributed about the true value, and show a similar spread to the unseeded case. All of the triggering parameters have much tighter posteriors. Even though some of the triggering parameter posteriors do not contain the true value, the percentage error remains small. This is due to the stochastic variability of these catalogues and this bias should decrease for longer catalogues. There is always trade-off between $\alpha$ and $K$ which is difficult to resolve. $K$ describes the magnitude independent productivity seen in the Omori law and $\alpha$ describes how the full triggering function productivity varies with magnitude; consequently one requires many sequences from parents of different magnitudes to resolve $K$ and $\alpha$ well.

Studies which are seeking to assign a physical cause to spatial and/or temporal variability of the background and triggering parameters should ensure that the variability cannot be explained by the stochastic nature of finite earthquake catalogues. The methods presented here provide one possible tool for doing this.

\begin{figure}[h!]
\begin{center}
\includegraphics[width=15cm]{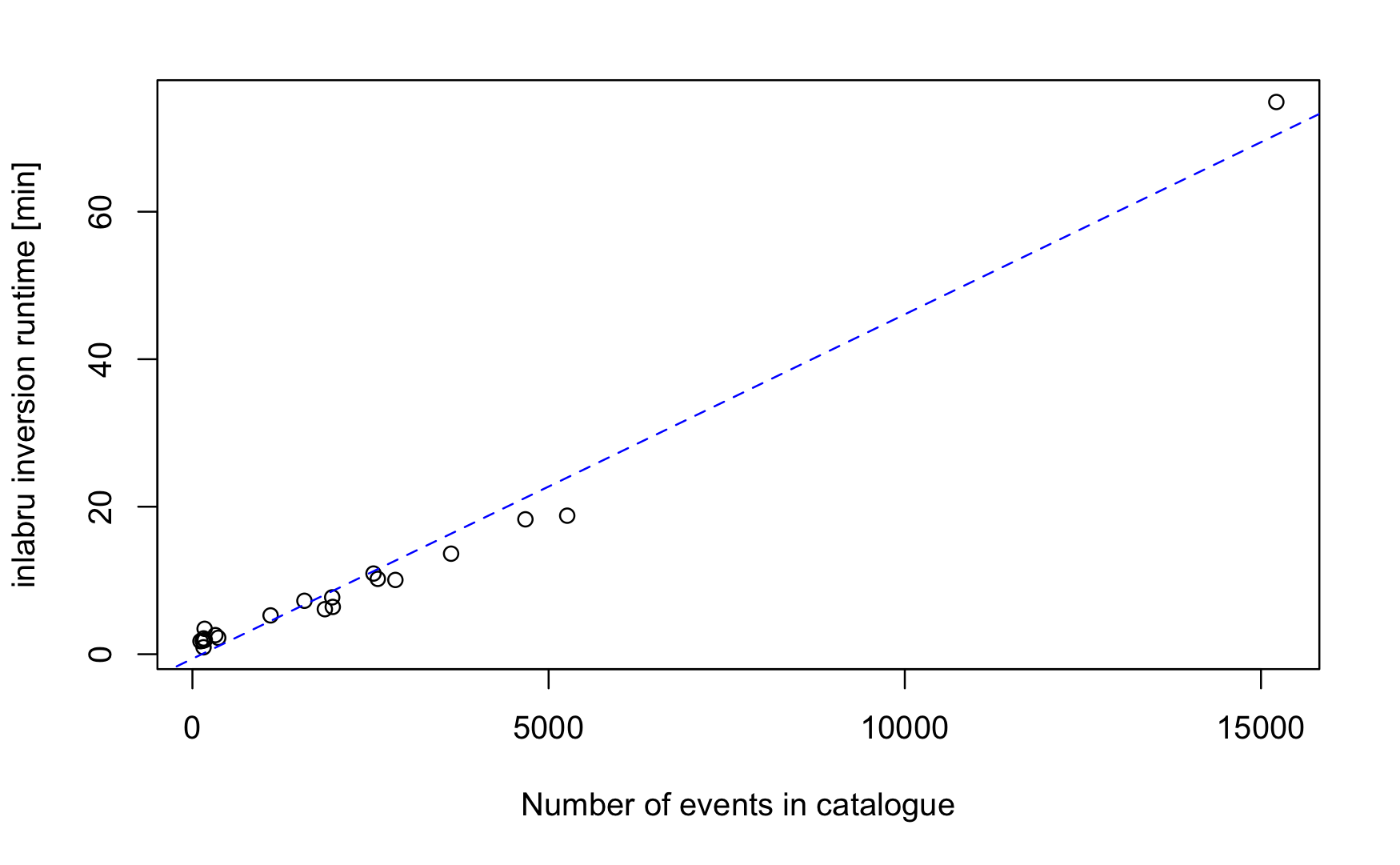}
\end{center}
\caption{The 20x1000 day synthetic catalogues presented in Figs.\ref{fig:tenQuietCats} and \ref{fig:10M6p7Cats} all have different numbers of events in them because of their stochastic nature. This figure plots the time taken for \inlabru to invert each of these catalogues as a function of the number of events in the catalogues. }\label{fig:compComplexity}
\end{figure}

Each of the 20 stochastic catalogues generated for this section have a different number of events. We timed the runtime for each analysis and have plotted it in Fig.\ref{fig:compComplexity} as a function of the number of events in the training catalogue. We find that not only is our \inlabru method ~10 times faster than "\bayesianETAS" for catalogues of more than 2500 events - but also that it scales relatively linearly with the number of events. We inverted a catalogue with ~15000 events in 70 minutes and it is likely this can be speeded up further using the the high performance sparse matrix solver \texttt{pardiso} package.

The inversions of synthetic data presented here show that the stochastic variability in the training catalogues produces understandable variability in the posteriors. More data and sequences containing both triggered sequences and background allow us to resolve all parameters well. Better resolution of $\alpha$ and $K$ would require aftershock sequences from parents of different sizes. We see that only having lower magnitude events leads to broad posteriors on the triggering parameters. The following section explore the impact of reducing the amount of background data on the resolution of $\mu$ for the seeded sequences.

\subsection{Importance of a representative sample} \label{representative}

The motivation for applying the ETAS model is sometimes the presence of an 'interesting' feature, such as an evolving or complex aftershock sequence following a notable event. In this section we explore whether it is important to have both quiet periods as well as the aftershock sequence itself for accurately recovering the true parameterisation. This motivates defining what a representative sample looks like; evaluating this in practice is non-trivial, but we can outline what is insufficient.

\begin{figure}[h!]
\begin{center}
\includegraphics[width=15cm]{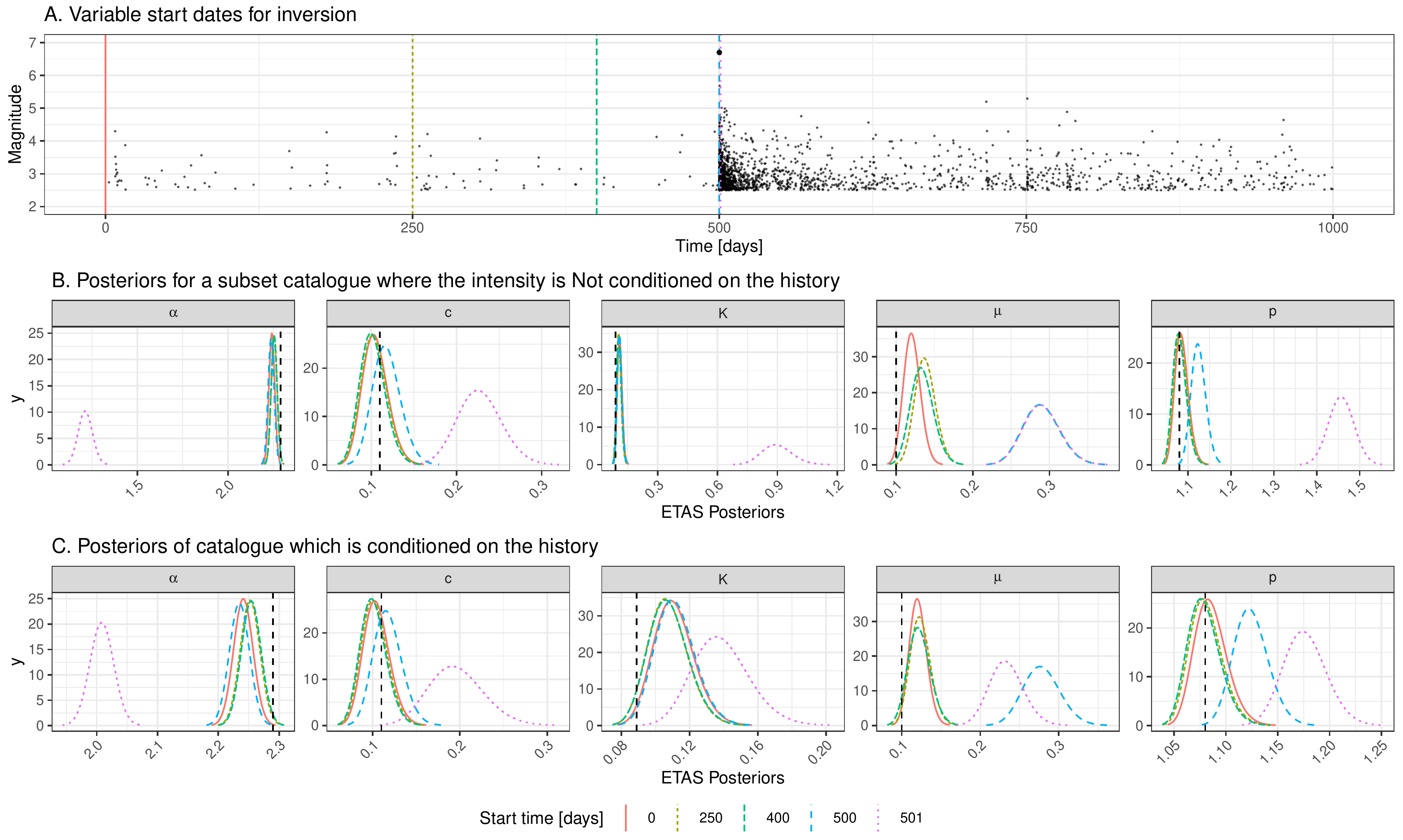}
\end{center}
\caption{(A) Catalogue used to explore the concept of a representative sample and history conditioning. The baseline case on the top row has 500 days of background and a M6.7 event on day 500 with the sequence being recorded until day 1000. We vary the start date for the analysis to remove the first 250 days, 400 days, 500 days and 501 days. In the last 2 cases there is no background period remaining and the large event is also prior to the catalogue subset for the final case. (B) Posteriors of the ETAS parameters for each of the catalogue subsets when we crop the sub catalogue and do not use the preceding data to condition the model. (C) Rather than cropping out sub-catalogues, we now retain the events preceding the start of the model domain and use these when estimating the triggering function. This produces a notably improved performance for the start date on day 501 when the large event is no-longer within the model domain.}\label{fig:varyRunin_post}
\end{figure}

We start with the a 1000 day catalogue including a M6.7 event seeded on day 500. We then generate catalogue subsets by eliminating the first 250, 400, 500 and 501 days of the catalogue (Start dates of subcatalogues shown as vertical dashed lines in Fig.\ref{fig:varyRunin_post}(A)) and rerun the \inlabru ETAS inversion on these subsets. Since the initial period is relatively quiet, we do not remove a large proportion of the events - however, we are removing events from the period where the background events are relatively uncontaminated by triggered events. In doing this, we explore what the necessary data requirements are for us to expect that \inlabru can reliably estimate both the background and triggering parameters. When we remove 501 days, we are also removing the seeded mainshock from the subcatalogue.

 First, we consider what happens to the posterior of the background rate, $\mu$, as the length of the sub-catalogues is shortened. With 500 days of background before the seeded event, we resolve $\mu$ accurately. As the quiet background is progressively removed, the model estimate of $\mu$ systematically rises. When there is between 250-100 days of background data, the mode overestimates the data-generating parameter by around 30\% but it still lies within the posterior distributions (turquoise and brown curves for $\mu$ in Fig.\ref{fig:varyRunin_post}(B)). When there is no background period, the overestimation of $\mu$ (blue curve for $\mu$ in Fig.\ref{fig:varyRunin_post}(B)) is on the order of a factor of 2.8. The estimate corresponds to the level of seismicity at the end of the model domain which has not decayed back to the background rate. From an operational perspective, it is much easier to extend the start date of training data back before the sequence of interest started than to wait until the background rate has been recovered. We should therefore expect an analysis looking for time varying background rate during the sequences carries a risk of bias by this effect.

 All of the models, apart from the one starting on day 501, contain the M6.7 event and 500 days of its aftershocks (Fig.\ref{fig:varyRunin_post}A). In these cases, the triggering parameters are well described by the posteriors  (Fig.\ref{fig:varyRunin_post}B). However, where the model domain starts on day 501 we lose the M6.7 event and its aftershocks on the first day. This results in significant bias in all the triggering parameters as well as the background rate  (Pink curve in Fig.\ref{fig:varyRunin_post}B). Modelling of specific sequences needs particular care to be taken in the choice of model domain and exclusion of the mainshock from the analysis can pose a major problem in conditioning the ETAS parameters.
 
The results already presented in Fig.\ref{fig:post_stochastic} showed that the inversion scheme struggles to recover the triggering parameters when there is no significant sequence in the dataset. Combined with the results for having no background period in this section, we argue that a representative sample should include periods of activity and inactivity if both the background rate and triggering parameters are to be estimated reliably. We also suggest mainshocks of different magnitudes would help for resolving $\alpha$. By running synthetics such as the ones presented here, one can gain insight into the data requirements in specific case studies.

The model where the mainshock was not part of the subcatalogue was particularly biased (pink curve in Fig.\ref{fig:varyRunin_post}B). In the next section, we explore whether we can correct for this by including the triggering effects of events that occurred prior to temporal domain being evaluated.

\subsection{Impact of historic run-in period} \label{historic}

In the previous section, we explicitly cropped out subcatalogues and ran the analysis on that subset of the data, effectively throwing the rest away. We demonstrated the consequences removing the M6.7 mainshock from the sub-catalogue being analysed; the posteriors on the triggering parameters and background became significantly biased (Fig.\ref{fig:varyRunin_post}(B)). This example talks to the wider need for the intensity function to be conditioned on historic events prior to the start of the model domain. This is a common issue in modelling regions that have experienced the largest earthquakes.

 In \texttt{ETAS.inlabru}, we have another option when analysing catalogue subsets. Rather than cropping out the data, we can provide an extended catalogue and specify a model domain that is smaller than the whole dataset. For the time component, this means that event prior to the temporal domain will have their triggering contribution to the model domain taken into account. This approach crops off events beyond the temporal model domain because they do not have a causal impact on the results. The model domain is defined using `T1` and `T2` in the input list for the start and end time respectively. \inlabru assumes any events in the catalogue prior to `T1` should be used to historic preconditioning.

In practice, this complicates the implementation of the time binning because events occurring prior to the start of the model domain only need to be evaluated from `T1` onward. The breaking of similarity of the time bins has a penalty in the speed of the implementation.

The results of conditioning the inversion using the historic events can be seen in Fig.\ref{fig:varyRunin_post}(C) and should be compared to the equivalent results for the cases where the subcatalogues did not have this preconditioning (Fig.\ref{fig:varyRunin_post}(B)). As the start date increases, the inclusion of small background events in the history has little effect on the results because their triggering effect is small. However, a significant improvement in the estimated posteriors is seen when the M6.7 mainshock is removed from the model domain (c.f pink lines for each parameter in Fig.\ref{fig:varyRunin_post}(B) and (C)). The historic pre-conditioning improves the estimation of all the triggering parameters when the mainshock is missing.

In the analysis of real catalogues, this effect will be particularly relevant when there have been very large past earthquakes which are still influencing today's rates.

\subsection{Impact of short term incompleteness}

Finally, we explore the effect of short term incompleteness after large mainshocks on the inverted parameters. This rate dependent incompleteness occurs because is hard to resolve the waveforms of small earthquakes when overprinted by many larger events, yet the effect is short lived. 

We take a 1000 day catalogue with a M6.7 event seeded on day 500 and then introduce a temporary increase in the completeness threshold after the M6.7 event using the functional form suggested by \cite{helmstetterIncompleteness2006},

\begin{equation}
    M_c(t)=M_i-G-H \log_{10}(t-t_i)
\end{equation}

where, $M_i$ and $t_i$ are the magnitude and occurrence time of the event we are modelling the incompleteness for, $t$ is the time we wish to evaluate the new completeness threshold for and $G$, $H$ are parameters of the model. We do not address here how these parameters should be determined in a real dataset and, informed by \cite{elstBPositive2021}, we set them to 3.8 and 1.0 respectively for our synthetic study. Further, in this exploratory analysis we do not include incompleteness effects for other events in the sequence - so it should be considered a relatively conservative analysis.

\begin{figure}[h!]
\begin{center}
\includegraphics[width=15cm]{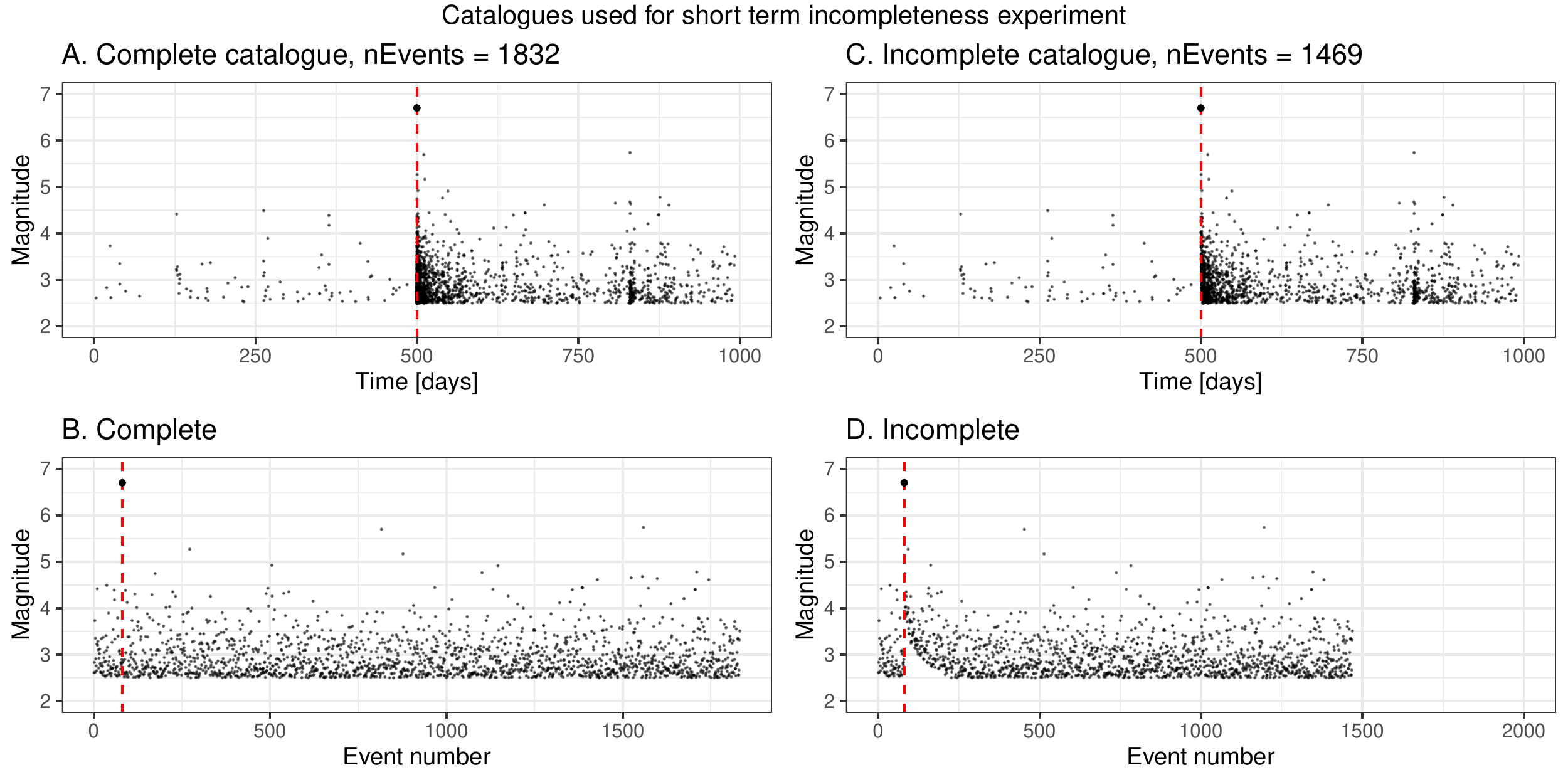}
\end{center}
\caption{Plots of the complete baseline catalogue and the catalogue with incompleteness artificially introduced using the functional form suggested by \cite{helmstetterIncompleteness2006}. The complete catalogue contains 1832 events and the incomplete has 1469 events. The time magnitude plot does not present this incompleteness well because it occurs in the very short term after the M6.7 event. The plot of magnitude as a function of event number clearly highlights the temporally varying incompleteness just after the large event.}\label{fig:incomplete_cat}
\end{figure}

\begin{figure}[h!]
\begin{center}
\includegraphics[width=15cm]{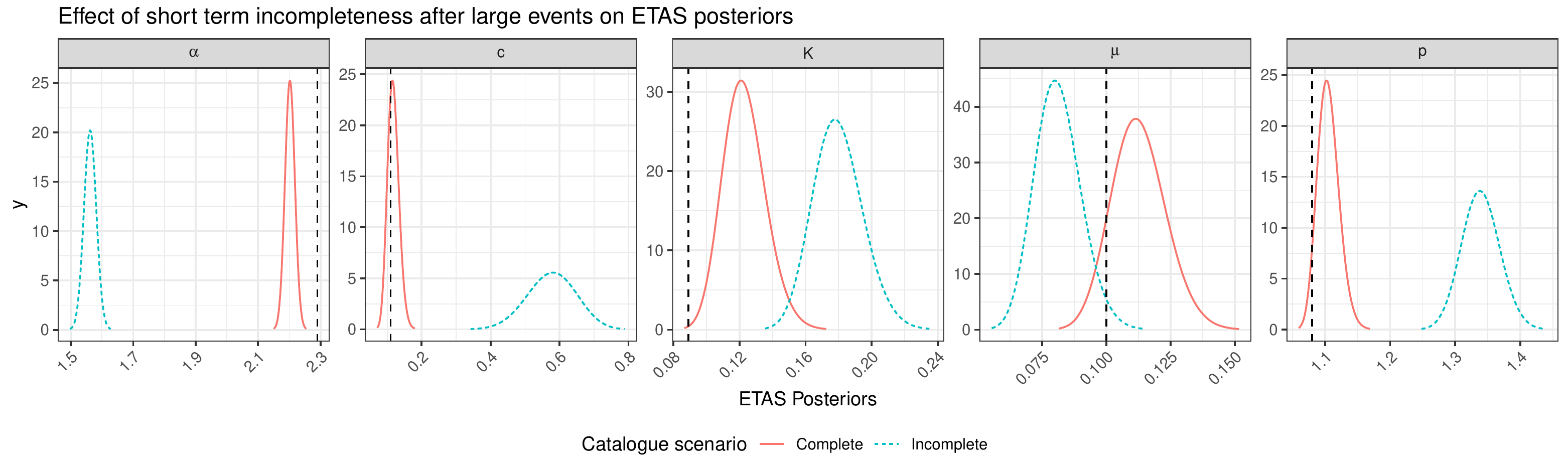}
\end{center}
\caption{Plot comparing the posteriors of the complete and incomplete catalogues presented in Fig.\ref{fig:incomplete_cat}. The true parameters are shown with the black dashed lines.} \label{fig:incompolete_post}
\end{figure}

We perform the inversion on the original catalogue and the one where short term incompleteness has removed a number of events (Fig.\ref{fig:incomplete_cat}) and compare the posteriors (Fig.\ref{fig:incompolete_post}).

The complete catalogue contained 1832 events and the incomplete catalogue contains 1469 events (Fig.\ref{fig:incomplete_cat}). This is difficult to see on the event time plot as most of these event are very close in time to the mainshock, so we have also plotted the magnitudes as a function of event number; here we see that after the mainshock (red dashed line) there is a transient in the completeness threshold.

\begin{figure}[h!]
\begin{center}
\includegraphics[width=15cm]{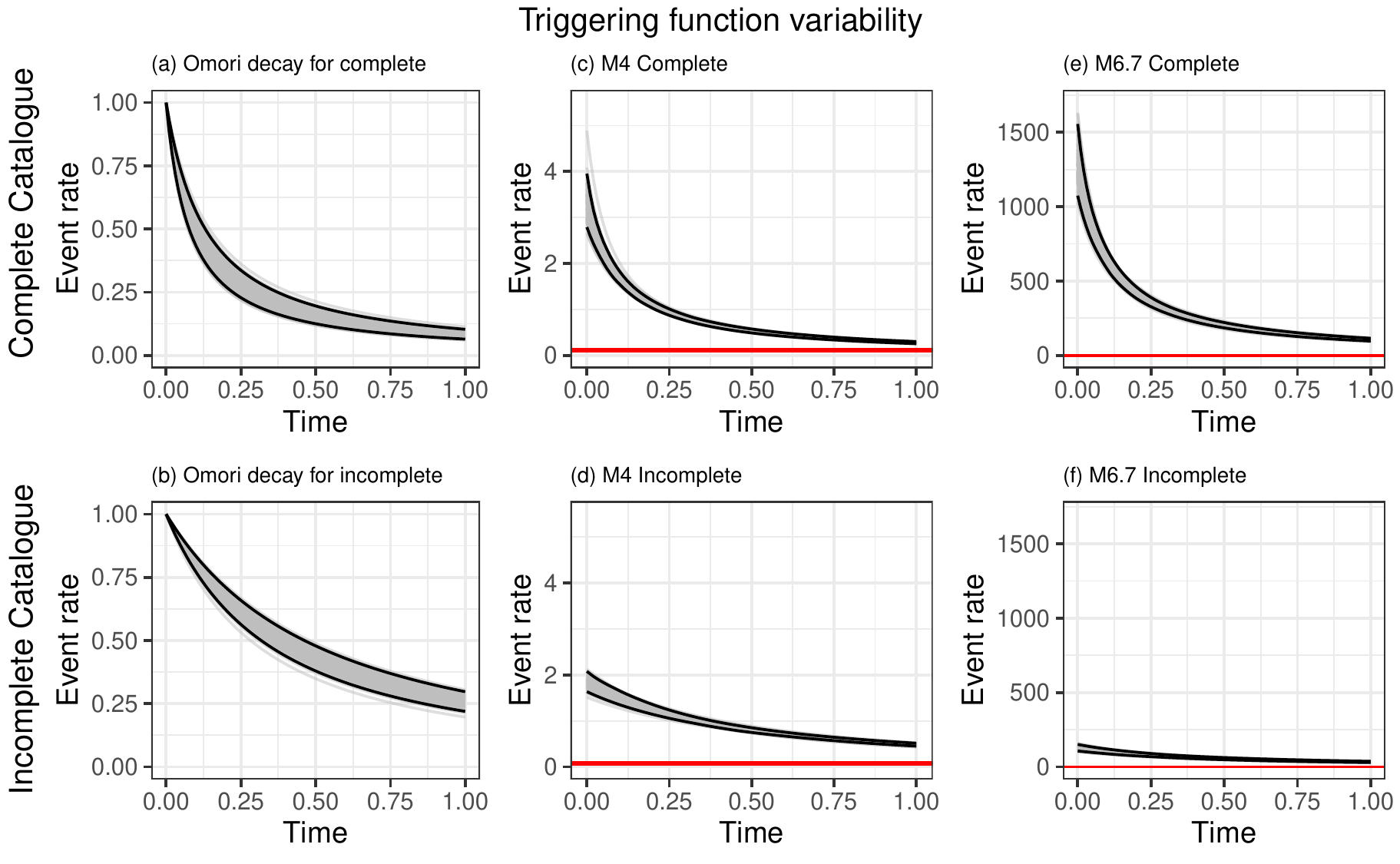}
\end{center}
\caption{Propagation of ETAS parameter uncertainty on the triggering functions. We take 100 samples of the ETAS posteriors for the complete 1000 day catalogue with an M6.7 on day 500 (top row) and for the temporally incomplete version of this catalogue as described in the text. We then use these samples to explore variability in the Omori decay (left hand column), the time-triggering function following an M4 event, and the  time-triggering function following an M7.6 event.}\label{fig:triggering_incomplete}
\end{figure}

All of the parameter estimates in the incomplete catalogue are now notably more biased and their standard deviation has not increased to compensate for this so the true values lie significantly outwith the posteriors (Fig.\ref{fig:incompolete_post}), and are therefore biased. The incomplete catalogue underestimates the background rate as there are fewer events in the same time period. Propagating the triggering parameter posteriors through to compare the triggering functions, we see extremely different triggering behaviour (Fig.\ref{fig:triggering_incomplete}). The Omori decay for the incomplete catalogue is longer lasting but the productivity, driven by $\alpha$ and $K$, is systematically lower.

The bias in the predicted triggering functions arising from short term incompleteness is significant and cannot be ignored within and OEF context. Solving this problem within \inlabru is beyond the scope of this paper.

\section{Discussion}

Having analysed a range of synthetic datasets, we now emphasise the lessons learned we should carry forward for the analysis of real sequences. In both examples below, the completeness is often assumed to be 2.5 and this compares well with our baseline synthetics.

Reliable inversions can only result from data that is representative of the processes the model is trying to capture. This means that datasets need to contain both productive sequences and periods that resolve the background without being overprinted by triggered events. The main difference between $\alpha$ and $K$ is that the former describes how the productivity varies with the magnitude of the triggering event whilst the latter is a magnitude insensitive productivity. If we are to resolve these uniquely, the training data would need relatively isolated sequences that are triggered by mainshocks of different ; this will be challenging in many use cases. 

Interpreting the results of an ETAS inversion is non-trivial. We advocate the routine analysis of synthetic to understand what it is possible to resolve in principle. 

Preconditioning models using large historic events can be significant. By considering samples of the triggering functions once can pre-determine the magnitude of events that need to be included as a function of time.

The use of synthetic modelling should be particularly important if time varying background rates are being inferred from the inversion of catalogue subsets in moving windows using the ETAS model.

The impact of short term incompleteness following large events is very significant and needs to be addressed either by raising the magnitude of completeness or formal modelling of the incompleteness. Resolving this for \texttt{ETAS.inlabru} is beyond the scope of this paper.

These are some of the considerations we explored here, but different use cases will present other modelling challenges that can be effectively explored through similar suits of synthetic modelling.

\section{Conclusions}

\texttt{ETAS.inlabru} is a fast and reliable tool for approximate Bayesian inference of the temporal ETAS model. The advantage of INLA over MCMC-based methods is that it is much faster. For large models, INLA finds a solution where MCMC methods would take far too long. For smaller problems, the speed of computation allows us to take a more exploratory and interactive approach to model construction and testing (\cite{wangINLA2018}). 

The exploratory approach illustrated here can be used to identify and understand sources of uncertainty and bias in the ETAS parameter posteriors that are derived from the structural and stochastic nature of the training data. We identify the need for a representative sample to contain periods of relative quiescence as well as sequences with clear triggering behaviour if all parameters are to be well resolved. 

Where studies focus solely on active sequences, the background rate can be erroneously estimated to be several times larger than the real background rate and the triggering parameters erroneously imply more rapidly decaying sequences than the true underlying parameterisation would. This implies that caution is needed in studies that allow the parameters to vary in time using windowing methods. Whist one cannot conclusively rule out that background rates and triggering behaviours may vary, we advocate that a stationary model with constant parameters should be adopted unless there is compelling evidence independent of the ETAS inversion, Colfiorito being a case in point (e.g. \cite{touatiColfiorito2014}).

Rate dependent incompleteness severely degrades the accuracy of the ETAS inversion and needs to be addressed directly.

The use of synthetic modelling, as presented here, provides a basis for discriminating when variations in the posteriors of ETAS parameters can be explained by deficiencies in the training data and when there is likely a robust and potentially useful signal. Such exploration requires a fast method for performing the inversion. The interpretation is easier when full posteriors can be compared, as opposed to just having point estimates. \inlabru is particularly well suited to this task.

\newpage

\section*{Conflict of Interest Statement}

The authors declare that the research was conducted in the absence of any commercial or financial relationships that could be construed as a potential conflict of interest.

\section*{Author Contributions}

The writing of this article and the analyses presented in this paper were led by Naylor. The ETAS model was implemented within \inlabru by Serafini. Lindgren provided advice on the implementation of the ETAS model using \inlabru.

\section*{Funding}
This work was founded by the Real-time Earthquake Risk Reduction for a Resilient Europe “RISE” project, which has received funding from the European Union's Horizon 2020 research and innovation program under grant Agreement 821115. 

\section*{Acknowledgments}
All the code to produce the present results is written in the R programming language. We have used the package $\texttt{ggplot2}$ (\cite{ggplot}). 

\section*{Supplemental Data}
 \href{http://home.frontiersin.org/about/author-guidelines#SupplementaryMaterial}{Supplementary Material} should be uploaded separately on submission, if there are Supplementary Figures, please include the caption in the same file as the figure. LaTeX Supplementary Material templates can be found in the Frontiers LaTeX folder.

\section*{Data Availability Statement}

The datasets [GENERATED/ANALYZED] for this study can be found in the [NAME OF REPOSITORY] [LINK].

We have generated an R-Package `\texttt{ETAS.inlabru}` and will be made available through CRAN. The results presented in this paper are reproducible using this package and a series of Rmd notebooks downloadeable from... It is intended that these are a good starting point for new users.

We have created a series of R Markdown notebooks to support this paper which can be found XX. There is a separate folder for each of the analyses in this paper. These are designed to enable the reader to start playing and modifying the examples presented here. It is our intention to create an R-Package that will streamline the implementation of the code, so the reader should view the code in the notebooks as a development version.

%Bibliography
\bibliographystyle{unsrt}  
\bibliography{references}

\begin{thebibliography}{10}

\bibitem{ogata1988etas}
Yosihiko Ogata.
\newblock Statistical models for earthquake occurrences and residual analysis
  for point processes.
\newblock {\em Journal of the American Statistical association}, 83(401):9--27,
  1988.

\bibitem{ogata2006space}
Yosihiko Ogata and Jiancang Zhuang.
\newblock Space--time etas models and an improved extension.
\newblock {\em Tectonophysics}, 413(1-2):13--23, 2006.

\bibitem{ogata2011significant}
Yosihiko Ogata.
\newblock Significant improvements of the space-time etas model for forecasting
  of accurate baseline seismicity.
\newblock {\em Earth, planets and space}, 63(3):217--229, 2011.

\bibitem{hawkes1971spectra}
Alan~G Hawkes.
\newblock Spectra of some self-exciting and mutually exciting point processes.
\newblock {\em Biometrika}, 58(1):83--90, 1971.

\bibitem{JSSv088c01}
Abdollah Jalilian.
\newblock Etas: An r package for fitting the space-time etas model to
  earthquake data.
\newblock {\em Journal of Statistical Software, Code Snippets}, 88(1):1–39,
  2019.

\bibitem{ross2021bayesian}
Gordon~J Ross.
\newblock Bayesian estimation of the etas model for earthquake occurrences.
\newblock {\em Bulletin of the Seismological Society of America},
  111(3):1473--1480, 2021.

\bibitem{rue2017bayesian}
H{\aa}vard Rue, Andrea Riebler, Sigrunn~H S{\o}rbye, Janine~B Illian, Daniel~P
  Simpson, and Finn~K Lindgren.
\newblock Bayesian computing with inla: a review.
\newblock {\em Annual Review of Statistics and Its Application}, 4:395--421,
  2017.

\bibitem{bachl2019inlabru}
Fabian~E Bachl, Finn Lindgren, David~L Borchers, and Janine~B Illian.
\newblock inlabru: an r package for bayesian spatial modelling from ecological
  survey data.
\newblock {\em Methods in Ecology and Evolution}, 10(6):760--766, 2019.

\bibitem{bayliss2020data}
Kirsty Bayliss, Mark Naylor, Janine Illian, and Ian~G Main.
\newblock Data-driven optimization of seismicity models using diverse data
  sets: Generation, evaluation, and ranking using inlabru.
\newblock {\em Journal of Geophysical Research: Solid Earth},
  125(11):e2020JB020226, 2020.

\bibitem{baylissCalifornia2022}
K.~Bayliss, M.~Naylor, F.~Kamranzad, and I.~Main.
\newblock Pseudo-prospective testing of 5-year earthquake forecasts for
  california using inlabru.
\newblock {\em Natural Hazards and Earth System Sciences}, 22(10):3231--3246,
  2022.

\bibitem{forlani2020joint}
Chiara Forlani, Samir Bhatt, Michela Cameletti, Elias Krainski, and Marta
  Blangiardo.
\newblock A joint bayesian space--time model to integrate spatially misaligned
  air pollution data in r-inla.
\newblock {\em Environmetrics}, 31(8):e2644, 2020.

\bibitem{riebler2016intuitive}
Andrea Riebler, Sigrunn~H S{\o}rbye, Daniel Simpson, and H{\aa}vard Rue.
\newblock An intuitive bayesian spatial model for disease mapping that accounts
  for scaling.
\newblock {\em Statistical methods in medical research}, 25(4):1145--1165,
  2016.

\bibitem{santermans2016spatiotemporal}
Eva Santermans, Emmanuel Robesyn, Tapiwa Ganyani, Bertrand Sudre, Christel
  Faes, Chantal Quinten, Wim Van~Bortel, Tom Haber, Thomas Kovac, Frank
  Van~Reeth, et~al.
\newblock Spatiotemporal evolution of ebola virus disease at sub-national level
  during the 2014 west africa epidemic: model scrutiny and data meagreness.
\newblock {\em PloS one}, 11(1):e0147172, 2016.

\bibitem{schrodle2011primer}
Birgit Schr{\"o}dle and Leonhard Held.
\newblock A primer on disease mapping and ecological regression using
  $\texttt{INLA}$.
\newblock {\em Computational statistics}, 26(2):241--258, 2011.

\bibitem{schrodle2011spatio}
Birgit Schr{\"o}dle and Leonhard Held.
\newblock Spatio-temporal disease mapping using inla.
\newblock {\em Environmetrics}, 22(6):725--734, 2011.

\bibitem{opitz2016extensive}
Nina Opitz, Caroline Marcon, Anja Paschold, Waqas~Ahmed Malik, Andrew Lithio,
  Ronny Brandt, Hans-Peter Piepho, Dan Nettleton, and Frank Hochholdinger.
\newblock Extensive tissue-specific transcriptomic plasticity in maize primary
  roots upon water deficit.
\newblock {\em Journal of Experimental Botany}, 67(4):1095--1107, 2016.

\bibitem{halonen2015road}
Jaana~I Halonen, Anna~L Hansell, John Gulliver, David Morley, Marta Blangiardo,
  Daniela Fecht, Mireille~B Toledano, Sean~D Beevers, Hugh~Ross Anderson,
  Frank~J Kelly, et~al.
\newblock Road traffic noise is associated with increased cardiovascular
  morbidity and mortality and all-cause mortality in london.
\newblock {\em European heart journal}, 36(39):2653--2661, 2015.

\bibitem{roos2015modeling}
Natalia~C Roos, Adriana~R Carvalho, Priscila~FM Lopes, and M~Grazia Pennino.
\newblock Modeling sensitive parrotfish (labridae: Scarini) habitats along the
  brazilian coast.
\newblock {\em Marine Environmental Research}, 110:92--100, 2015.

\bibitem{teng2022bayesian}
Jiaqi Teng, Shuzhen Ding, Huiguo Zhang, Kai Wang, and Xijian Hu.
\newblock Bayesian spatiotemporal modelling analysis of hemorrhagic fever with
  renal syndrome outbreaks in china using r-inla.
\newblock {\em Zoonoses and Public Health}, 2022.

\bibitem{bakka2018spatial}
Haakon Bakka, H{\aa}vard Rue, Geir-Arne Fuglstad, Andrea Riebler, David Bolin,
  Janine Illian, Elias Krainski, Daniel Simpson, and Finn Lindgren.
\newblock Spatial modeling with r-inla: A review.
\newblock {\em Wiley Interdisciplinary Reviews: Computational Statistics},
  10(6):e1443, 2018.

\bibitem{blangiardo2013spatial}
Marta Blangiardo, Michela Cameletti, Gianluca Baio, and H{\aa}vard Rue.
\newblock Spatial and spatio-temporal models with r-inla.
\newblock {\em Spatial and spatio-temporal epidemiology}, 4:33--49, 2013.

\bibitem{gomez2020bayesian}
Virgilio G{\'o}mez-Rubio.
\newblock {\em Bayesian inference with INLA}.
\newblock CRC Press, 2020.

\bibitem{taylor2014inla}
Benjamin~M Taylor and Peter~J Diggle.
\newblock Inla or mcmc? a tutorial and comparative evaluation for spatial
  prediction in log-gaussian cox processes.
\newblock {\em Journal of Statistical Computation and Simulation},
  84(10):2266--2284, 2014.

\bibitem{serafini2022approximation}
Francesco Serafini, Finn Lindgren, and Mark Naylor.
\newblock Approximation of bayesian hawkes process models with inlabru.
\newblock {\em arXiv preprint arXiv:2206.13360}, 2022.

\bibitem{hawkes1971point}
Alan~G Hawkes.
\newblock Point spectra of some mutually exciting point processes.
\newblock {\em Journal of the Royal Statistical Society: Series B
  (Methodological)}, 33(3):438--443, 1971.

\bibitem{touati2009origin}
Sarah Touati, Mark Naylor, and Ian~G Main.
\newblock Origin and nonuniversality of the earthquake interevent time
  distribution.
\newblock {\em Physical Review Letters}, 102(16):168501, 2009.

\bibitem{helmstetterIncompleteness2006}
Agnès Helmstetter, Yan~Y. Kagan, and David~D. Jackson.
\newblock {Comparison of Short-Term and Time-Independent Earthquake Forecast
  Models for Southern California}.
\newblock {\em Bulletin of the Seismological Society of America},
  96(1):90--106, 2006.

\bibitem{elstBPositive2021}
Nicholas~J. van~der Elst.
\newblock B-positive: A robust estimator of aftershock magnitude distribution
  in transiently incomplete catalogs.
\newblock {\em Journal of Geophysical Research: Solid Earth},
  126(2):e2020JB021027, 2021.

\bibitem{wangINLA2018}
Wang, Yu~Yue, and Faraway.
\newblock {\em Bayesian Regression Modeling with INLA}.
\newblock Chapman \& Hall, UK United Kingdom, 2018.

\bibitem{touatiColfiorito2014}
Sarah Touati, Mark Naylor, and Ian~G. Main.
\newblock {Statistical Modeling of the 1997–1998 Colfiorito Earthquake
  Sequence: Locating a Stationary Solution within Parameter Uncertainty}.
\newblock {\em Bulletin of the Seismological Society of America},
  104(2):885--897, 2014.

\bibitem{ggplot}
Hadley Wickham.
\newblock {\em ggplot2: Elegant Graphics for Data Analysis}.
\newblock Springer-Verlag New York, 2016.

\end{thebibliography}

\end{document}